\documentstyle[prd,aps,epsf,floats]{revtex}

\newcommand{\lsim}{\mathrel{\lower4pt\hbox{$\sim$}}
\hskip-12.5pt\raise1.6pt\hbox{$<$}\;}

\newcommand{\gsim}{\mathrel{\lower4pt\hbox{$\sim$}}
\hskip-12.5pt\raise1.6pt\hbox{$>$}\;}

\newcommand{\real}{{\rm Re}\,}
\newcommand{\imag}{{\rm Im}\,}
\newcommand{\sss}{\scriptscriptstyle}
\def\fH{{\cal H}}

\newcommand{\be}{\begin{equation}}
\newcommand{\ee}{\end{equation}}
\newcommand{\bea}{\begin{eqnarray}}
\newcommand{\eea}{\end{eqnarray}}
\newcommand{\ba}{\begin{array}}
\newcommand{\ea}{\end{array}}

\newcommand{\np}{Nucl.\ Phys.}

\newcommand{\zp}{Zeit.\ Phys.} 

\newcommand{\etal}{{\it et al}.,\ }

\begin{document}
\draft

\title{Phenomenology of two Higgs doublet models with flavor 
changing neutral currents}
\author{David Atwood}
\address{Theory Group, Continuous Electron Beam Accelerator Facility, 
Newport News, VA 23606}
\author{Laura Reina}
\address{Physics Department, Brookhaven National Laboratory, Upton, NY
11973}
\author{Amarjit Soni}
\address{Physics Department, Brookhaven National Laboratory, Upton, NY
11973}

\maketitle
\begin{abstract}
A comprehensive phenomenological analysis of a two Higgs doublet
model, with flavor-changing scalar currents at the tree-level, called
model~III, is presented. Constraints from existing experimental
information especially on $\Delta F=2$ processes are systematically
incorporated. Constraints emerging from rare $B$-decays, $Z\rightarrow
b\bar b$, and the $\rho$-parameter are also examined. Experimental
implications for $e^+e^- (\mu^+\mu^-)\rightarrow t\bar c+\bar t c$,
$t\rightarrow c\gamma(Z,g)$, $D^0$-$\bar D^0$, and $B^0_s$-$\bar
B^0_s$ oscillations, and for $e^+e^-(Z)\rightarrow b\bar s+\bar b s$
are investigated and experimental effort towards these is stressed. We
also emphasize the importance of clarifying the experimental issues
pertaining to $Z\rightarrow b\bar b$.
\end{abstract} 
\pacs{13.65.+i,12.60.Fr,11.30.Hv,14.65.Ha,14.80.Cp}

\section{Introduction}
\label{intro}

Recently we have been examining various issues \cite{eetc,mumutc}, in
a class of two Higgs doublet models (2HDM's) which allow flavor
changing neutral currents (FCNC's) at the tree level
\cite{sher,antaramian,hall,savage,lukesavage}.  In this work we want
to present a comprehensive analysis which gives the details of the
analytical calculations we used and summarize the status of our
knowledge regarding this type of 2HDM's\null. In particular we will
examine the important constraints and derive quantitative bounds on
the mass parameters and flavor changing (FC) couplings of the new
scalar fields based on existing low energy experiments.

FCNC's are naturally suppressed in the standard model (SM) because
they are forbidden at the tree level. However, at the one loop level,
the amplitudes for the Feynman diagrams which generate FC processes
tend to increase with virtual quark masses. Due to the large disparity
in masses of the up-type quarks, the GIM \cite{gim} suppression gets
very effectively removed in loops of these (virtual) quarks.
Therefore, FC transitions involving a pair of down-type quarks get
enhanced, due primarily to the presence of a top-quark in the loop,
while FC transitions which involve up-type quarks are usually very
small. This motivates the interest in processes like $b\rightarrow
s,d\gamma$ instead of $t\rightarrow c\gamma,Z$ or for $K^0\!-\!\bar
K^0$ and $B^0\!-\!\bar B^0$ mixing instead of $D^0-\bar D^0$ mixing.

There are many ways in which the extensions of the SM lead to FC
couplings at the tree level. For instance, as soon as we go from a
theory with one doublet of scalar fields to a theory with two
doublets, FCNC's are generated in the scalar sector of the theory. In
this case, couplings such as $tc\gamma$ or $tcZ$ can get enhanced too,
in much the same manner as there is an enhancement of $bs\gamma$ in
the SM\null. The interest in this class of implications is obvious: we
could have a clear signal of new physics, since the SM prediction for
any process involving a $tc\gamma$ or a $tcZ$ vertex is extremely
small.

As is well known, a model with tree level FCNC's will also have many
important repercussions for the $\Delta F=2$ mixing processes such as
$K^0-\bar K^0$, $B^0-\bar B^0$, and $D^0-\bar D^0$.  Indeed the
measured size of the mixing amplitude for $K^0-\bar K^0$, known for a
very long time, is so small that it places severe restrictions on the
FC sector of extended models.  This led Glashow and Weinberg
\cite{glash} to propose an {\it ad hoc\/} discrete symmetry whose sole
purpose was to forbid tree-level FCNC's to appear in models with more
than one Higgs doublet.  In particular, for the simplest case of
2HDM's, depending on whether the up-type and down-type quarks couple
to the same or to two different scalar doublets leads to two versions
of such models, called model~I and model~II\null.  Both of them share
with the SM the distinctive feature that they do not allow tree-level
FCNC's\null. These models have been extensively studied and a
multitude of interesting implications have been discussed in the
literature.

Our starting point for investigating tree-level FCNC's is primarily
based on the realization that the top quark may be quite different
from the lighter quarks.  It may well be that the theoretical
prejudice of the non existence of tree level FCNC's, based on
experiments involving the lighter quarks, is not relevant to the top
quark.  This leads one to formulate a model which allows the
possibility of large tree level FCNC's involving the top quark while
it keeps the FCNC's of the lighter quarks, especially those involving
the quarks of the first family, at a negligible level. A rather
natural way of implementing this notion is by taking the new scalar FC
vertices to be proportional to the masses of the participating quarks
at the vertex \cite{sher,antaramian,hall}.  A hierarchy is then
automatically introduced which enhances the top quark couplings while
keeping the couplings of the lighter quarks at some order of magnitude
smaller.  The compatibility of any such assumption with the many
constraints mentioned above clearly needs to be checked and the
allowed region of the parameter space determined, in particular,
scalar masses and couplings.  This type of 2HDM is now referred to as
model~III \cite{hou}.

Our first interest in model~III was motivated by the idea of looking
for top-charm production at an $e^+e^-$ and/or $\mu^+\mu^-$ collider
\cite{eetc,mumutc}. We were intrigued by the possibility of a clear
signal for the reaction $e^+e^-\rightarrow t\bar c$ for which the SM
prediction is extremely small. The reaction has a distinctive
kinematical signature in a very clean environment.  These
characteristics, which are unique to the lepton colliders, should
compensate for the lower statistics one expects compared to those at
hadron colliders.  In this paper we present the details of our
calculations for this important process.

After a brief overview of model~III in Sec.~\ref{model}, we will
discuss $e^+e^-\rightarrow t\bar c+\bar t c$ in Sec.~\ref{eetc} and
present some of the relevant formulae in detail in
Appendix~\ref{formfactors}\null.  In Sec.~\ref{tcgZ} we will consider
the rare decays $t\rightarrow c\gamma,Z$ and $g$; herein we will also
compare our results with those existing in the literature for
$t\rightarrow c\gamma$ and $cZ$ in model~III.  Top-charm production at
a $\mu^+\mu^-$ collider is briefly discussed in Sec.~\ref{mumutc}.  In
Sec.~\ref{mixing} we consider the repercussions of the $\Delta F=2$
mixing processes and extract the constraints that emerge on parameters
of model~III\null.  Sec.~\ref{bsg} discusses the impact of the
experimental results for $Br(B\rightarrow X_s\gamma)$, the
$\rho$-parameter and $Z\rightarrow b\bar b$.  We then examine in
Sec. \ref{b0s} the physical consequences of the constrained physical
model for the $B_s^0-\bar B_s^0$ oscillations, the flavor-changing $Z$
decay $Z\rightarrow \bar b s+ b\bar s$, and also some rare $B$-decays.
Sec.~\ref{end} offers the outlook and the conclusions.

\section{The model}
\label{model}

A mild extension of the SM with one additional scalar SU(2) doublet
opens up the possibility of flavor changing scalar currents (FCSC's)
at the tree level.  In fact, when the up-type quarks and the down-type
quarks are allowed simultaneously to couple to more than one scalar
doublet, the diagonalization of the up-type and down-type mass
matrices does not automatically ensure the diagonalization of the
couplings with each single scalar doublet. For this reason, the 2HDM
scalar potential and Yukawa Lagrangian are usually constrained by
an {\it ad hoc\/} discrete symmetry \cite{glash}, whose only role is
to protect the model from FCSC's at the tree level. Let us consider a
Yukawa Lagrangian of the form

\be
{\cal L}^{(III)}_{Y}= \eta^{U}_{ij} \bar Q_{i,L} \tilde\phi_1 U_{j,R} +
\eta^D_{ij} \bar Q_{i,L}\phi_1 D_{j,R} + 
\xi^{U}_{ij} \bar Q_{i,L}\tilde\phi_2 U_{j,R}
+\xi^D_{ij}\bar Q_{i,L} \phi_2 D_{j,R} \,+\, h.c. 
\label{lyukmod3}
\ee

\noindent where $\phi_i$, for $i=1,2$, are the two scalar doublets of
a 2HDM, while $\eta^{U,D}_{ij}$ and $\xi_{ij}^{U,D}$ are the 
non-diagonal matrices of the Yukawa couplings.  Imposing the following
{\it ad hoc\/} discrete symmetry

\bea
\phi_1 \rightarrow -\phi_1 \,\,\,\,\,\,\,\,\,\mbox{and}&&
\,\,\,\,\,\phi_2\rightarrow\phi_2\\ \label{discr_symm}
D_i\rightarrow -D_i \,\,\,\,\,\,\,\,\,\mbox{and}&&\,\,\,\,\, 
U_i\rightarrow\mp U_i\nonumber
\eea

\noindent one obtains the so called model~I and model~II, depending on
whether the up-type and down-type quarks are coupled to the same or to
two different scalar doublets respectively \cite{hunter}.

In contrast we will consider the case in which no discrete symmetry
is imposed and both up-type and down-type quarks then have FC
couplings. For this type of 2HDM, which we will call model~III, the
Yukawa Lagrangian for the quark fields is as in Eq.~(\ref{lyukmod3})
and no term can be dropped {\it a priori}, see also
refs.~\cite{lukesavage,eetc}.

For convenience we can choose to express $\phi_1$ and $\phi_2$ in a
suitable basis such that only the $\eta_{ij}^{U,D}$ couplings generate
the fermion masses, i.e.\ such that

\be
\langle\phi_1\rangle=\left( 
\begin{array}[]{c}
0\\
{v/\sqrt{2}}
\end{array}
\right)\,\,\,\, , \,\,\,\,
\langle\phi_2\rangle=0 \,\,\,.
\ee 

\noindent The two doublets are in this case of the form

\be
\phi_1=\frac{1}{\sqrt{2}}\left[\left(\ba{c} 0 \\ v+H^0\ea\right)+
\left(\ba{c} \sqrt{2}\,\chi^+\\ i\chi^0\ea\right)\right]\,\,\,\,;\,\,\,\,
\phi_2=\frac{1}{\sqrt{2}}\left(\ba{c}\sqrt{2}\,H^+\\ H^1+i H^2\ea\right)
\,\,\,.
\ee

\noindent The scalar Lagrangian in the ($H^0$, $H^1$, $H^2$,
$H^{\pm}$) basis is such that\cite{knowles,hunter}:

\begin{enumerate}

\item the doublet $\phi_1$ corresponds to the scalar doublet of the SM
and $H^0$ to the SM Higgs field (same couplings and no interactions
with $H^1$ and $H^2$);

\item all the new scalar fields belong to the $\phi_2$ doublet;

\item both $H^1$ and $H^2$ do not have couplings to the gauge bosons
of the form $H^{1,2}ZZ$ or $H^{1,2}W^+W^-$.

\end{enumerate}

\noindent However, while $H^{\pm}$ is also the charged scalar mass
eigenstate, ($H^0$, $H^1$, $H^2$) are not the neutral mass
eigenstates. Let us denote by ($\bar H^0$, $h^0$) and $A^0$ the two
scalar plus one pseudoscalar neutral mass eigenstates. They are
obtained from ($H^0$, $H^1$, $H^2$) as follows

\bea 
\label{masseigen}
\bar H^0 & = & \left[(H^0-v)\cos\alpha + H^1\sin\alpha \right]
\nonumber \\ 
h^0 & = & \left[-(H^0-v)\sin\alpha + H^1\cos\alpha \right] \\ 
A^0 & = &  H^2 \nonumber 
\eea 

\noindent where $\alpha$ is a mixing angle, such that for
$\alpha\!=\!0$, ($H^0$, $H^1$, $H^2$) coincide with the mass
eigenstates.  We find it more convenient to express $H^0$, $H^1$, and
$H^2$ as functions of the mass eigenstates, i.e.

\bea
\label{nomasseigen}
H^0 &=& \left(\bar H^0\cos\alpha-
h^0\sin\alpha\right)+v \nonumber \\
H^1 &=& \left( h^0\cos\alpha+\bar H^0\sin\alpha\right) \\
H^2 &=& A^0 \,\,\,.\nonumber
\eea

\noindent In this way we may take advantage of the mentioned
properties (1), (2), and (3), as far as the calculation of the
contribution from new physics goes. In particular, only the $\phi_1$
doublet and the $\eta^U_{ij}$ and $\eta^D_{ij}$ couplings are involved
in the generation of the fermion masses, while $\phi_2$ is responsible
for the new couplings.

After the rotation that diagonalizes the mass matrix of the quark
fields, the FC part of the Yukawa Lagrangian looks like

\be
{\cal L}_{Y,FC}^{(III)} = \hat\xi^{U}_{ij} \bar Q_{i,L}\tilde\phi_2 U_{j,R} 
 +\hat\xi^D_{ij}\bar Q_{i,L} \phi_2 D_{j,R} \,+\, h.c. \label{lyukfc}
\ee

\noindent where $Q_{i,L}$, $U_{j,R}$, and $D_{j,R}$ denote now the
quark mass eigenstates and $\hat\xi_{ij}^{U,D}$ are the rotated
couplings, in general not diagonal. If we define $V_{L,R}^{U,D}$ to be
the rotation matrices acting on the up- and down-type quarks, with
left or right chirality respectively, then the neutral FC couplings
will be

\be
\hat\xi^{U,D}_{\rm neutral}=(V_L^{U,D})^{-1}\cdot \xi^{U,D}
\cdot V_R^{U,D} \,\,\,.
\label{neutral}
\ee

\noindent On the other hand for the charged FC couplings we will have

\bea
\hat\xi^{U}_{\rm charged}\!&=&\!\hat\xi^{U}_{\rm neutral}\cdot 
V_{\sss{\rm CKM}}\nonumber\\
\hat\xi^{D}_{\rm charged}\!&=&\!V_{\sss{\rm CKM}}\cdot
\hat\xi^{D}_{\rm neutral} \label{charged}
\eea 

\noindent where $V_{\sss{\rm CKM}}$ denotes the
Cabibbo-Kobayashi-Maskawa matrix. To the extent that the definition of
the $\xi^{U,D}_{ij}$ couplings is arbitrary, we can take the rotated
couplings as the original ones. Thus, we will denote by
$\xi^{U,D}_{ij}$ the new rotated couplings in Eq.~(\ref{neutral}),
such that the charged couplings in Eq.~(\ref{charged}) look like
$\xi^{U}\cdot V_{\sss{\rm CKM}}$ and $V_{\sss{\rm CKM}}\cdot\xi^{D}$.
This form of the charged couplings is indeed peculiar to model~III:
they appear as a linear combination of neutral FC couplings multiplied
by some CKM matrix elements.  This is an important distinction between
model~III on the one hand, and models~I and II on the other. As we
will see in the phenomenological analysis this can have important
repercussions for many different physical quantities.

In order to apply to specific processes we have to make some definite
ansatz on the $\xi_{ij}^{U,D}$ couplings. Many different suggestions
can be found in the literature \cite{sher,antaramian,hall,eetc}. In
addition to symmetry arguments, there are also arguments based on the
widespread perception that these new FC couplings are likely to mainly
affect the physics of the third generation of quarks only, in order to
be consistent with the constraints coming from $K^0\!-\!\bar K^0$ and
$B^0\!-\!\bar B^0$. A natural hierarchy among the different quarks is
provided by their mass parameters, and that has led to the assumption
that the new FC couplings are proportional to the mass of the quarks
involved in the coupling. Most of these proposals are well described
by the following equation

\be
\xi^{U,D}_{ij}=\lambda_{ij}\,\frac{\sqrt{m_i m_j}}{v} \label{coupl_sher}
\ee 

\noindent which basically coincides with what was proposed by Cheng
and Sher \cite{sher}. In this ansatz the residual degree of
arbitrariness of the FC couplings is expressed through the
$\lambda_{ij}$ parameters, which need to be constrained by the
available phenomenology. In particular we will see how $K^0\!-\!\bar
K^0$ and $B^0\!-\!\bar B^0$ mixings (and to a less extent
$D^0\!-\!\bar D^0$ mixing) put severe constraints on the FC couplings
involving the first family of quarks. Additional constraints are given
by the combined analysis of the $Br(B\rightarrow X_s\gamma)$, the
$\rho$-parameter, and $R_b$, the ratio of the $Z\rightarrow b\bar b$
rate to the $Z$-hadronic rate. We will analyze all these constraints
in the following sections and discuss the resulting configuration of
model~III at the end of Sec.~\ref{bsg}.

\section{Top-charm production at $\lowercase{e}^+\lowercase{e}^-$ 
colliders}
\label{eetc}

The presence of FC couplings in the Yukawa Lagrangian of
Eq.~(\ref{lyukmod3}) affects the top-charm production at both hadron
and lepton colliders. In particular we want to study top-charm
production at lepton colliders, because, as we have emphasized before
\cite{eetc,mumutc}, in this environment the top-charm production has a
particularly clean and distinctive signature. In principle, the
production of top-charm pairs arises both at the tree level, via the
$s$-channel exchange of a scalar field with FC couplings, and at the
one loop level, via corrections to the $Ztc$ and $\gamma tc$
vertices. We will consider in this section the case of an $e^+e^-$
collider and in Sec.~\ref{mumutc} that of a $\mu^+\mu^-$ collider.

The $s$-channel top-charm production is one of the new interesting
possibilities offered by a $\mu^+\mu^-$ collider in studying the
physics of standard and non-standard scalar fields (see
Sec.~\ref{mumutc} and refs.~therein). However, it is not relevant for
an $e^+e^-$ collider, because the coupling of the scalar fields to the
electron is likely to be very suppressed (see
Eq.~(\ref{coupl_sher})). Therefore we will consider top-charm
production via $\gamma$ and $Z$ boson exchange, i.e.\ the process
$e^+e^-\rightarrow\gamma^*,Z^*\rightarrow \bar t c+\bar c t$, where
the effective one loop $\gamma tc$ or $Ztc$ vertices are induced by
scalars with FC couplings (e.g.\ model~III) \cite{model3only}.

Let us write the one loop effective vertices $Ztc$ and $\gamma tc$ in
the following way

\be
\Delta^{(V)}_{tc}= \frac{1}{16\pi^2}\bar c\left(A^{(V)}\gamma^{\mu}+
B^{(V)}\gamma^{\mu}\gamma^5+iC^{(V)}\sigma^{\mu\nu}
\frac{q_{\nu}}{m_t}+iD^{(V)}\sigma^{\mu\nu}\frac{q_{\nu}}{m_t}
\gamma^5\right)tV_{\mu}
\label{oneloopvertex}
\ee

\noindent where $V=\gamma,Z$ and $A^{(\gamma,Z)}$, $B^{(\gamma,Z)}$,
\ldots denote the form factors generated by one loop corrections. A
sample of the corrections one has to compute in model~III is given in
Fig.~1, for both the neutral and the charged scalar fields. The analytical
expressions are given in Appendix~\ref{formfactors} where the details
of the form factor calculation are also explained.

\begin{figure}
\centering
\epsfxsize=4.in
\leavevmode\epsffile{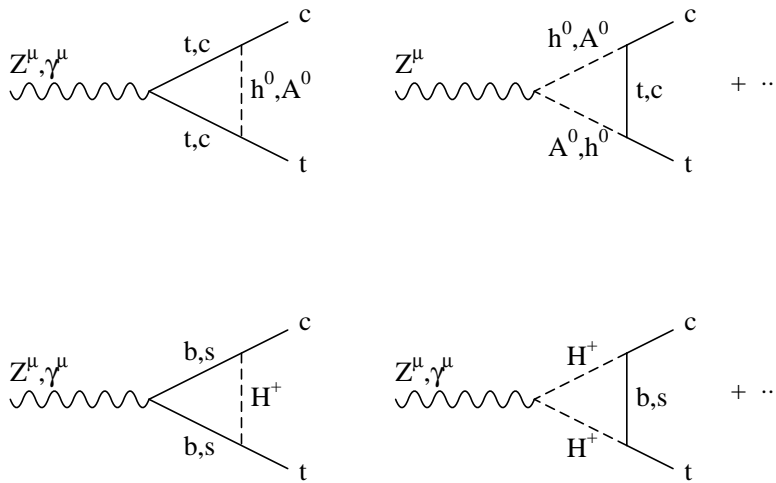}
\caption[]{Example of one-loop contributions to the $Ztc$ and $\gamma
tc$ vertices in model~III.}
\label{Zgtc_1loop}
\end{figure}

In terms of the $A^{(\gamma,Z)}$, $B^{(\gamma,Z)}$,\ldots form factors
we can compute the cross section for $e^+e^-\rightarrow
\gamma^*,Z^*\rightarrow\bar c t+\bar t c$. The total cross section
will be the sum of three terms

\be
\sigma_{tot}=\sigma_\gamma+\sigma_Z+\sigma_{\gamma Z}
\ee

\noindent corresponding to the pure photon, pure $Z$ and photon-$Z$
interference contributions respectively. Thus 

\be
\sigma_\gamma=\frac{1}{128\pi^4}
\frac{\alpha_e}{2s}N_c\beta^4\left(1-\frac{1}{3}\beta^2
\right)\left[|A^{\gamma}|^2+|B^\gamma|^2-\frac{1}{(1-\beta^2)}
(|C^\gamma|^2+|D^\gamma|^2)\right] \label{sigma_ph}
\ee

\be
\sigma_Z=\frac{1}{128\pi^4}
\frac{s}{D(s)^2}\frac{\alpha_e}{16s_{\sss W}^2c_{\sss W}^2}
N_c(1-4s_{\sss W}^2+8s_{\sss W}^4)\beta^4\left(1-\frac{1}{3}\beta^2\right)
\left[|A^Z|^2+|B^Z|^2-\frac{1}{(1-\beta^2)}(|C^Z|^2+|D^Z|^2)\right]
\ee

\bea
\sigma_{\gamma Z}&=&\frac{1}{128\pi^4}
\frac{\alpha_e}{4s_{\sss W}c_{\sss W}}N_c
(1-4s_{\sss W}^2)\beta^4\left(1-\frac{1}{3}\beta^2\right)\cdot\left\{
\phantom{\frac{1}{2}}\real D(s)\left[\phantom{\frac{1}{2}}\!\!
\real A^\gamma\real A^Z+\imag A^\gamma\imag A^Z+\real B^\gamma\real B^Z
\right.\right. \nonumber\\
&+&\left.\imag B^\gamma\imag B^Z-
\frac{1}{(1-\beta^2)}
\left(\real C^\gamma\real C^Z+\imag C^\gamma\imag C^Z+\real D^\gamma
\real D^Z+\imag D^\gamma\imag D^Z\right)\right]\nonumber\\
&+&\imag D(s)\left[\phantom{
\frac{1}{2}}\!\!\real A^\gamma\imag A^Z-\imag A^\gamma\real A^Z+
\real B^\gamma\imag B^Z-\imag B^\gamma\real B^Z\right.\\
&-&\left.\left.\frac{1}{(1-\beta^2)}
\left(\real C^\gamma\imag C^Z-\imag C^\gamma\real C^Z+\real D^\gamma
\imag D^Z-\imag D^\gamma\real D^Z\right)\right]\right\}\nonumber
\eea

\noindent where $\alpha_e$ denotes the QED fine structure constant,
$s_{\sss W}\!=\!\sin\theta_{\sss W}$, $c_{\sss W}\!=\!\cos\theta_{\sss
W}$, $s\!=\!q^2$ is the center of mass energy squared, $D(s)$ denotes
the $Z$ boson propagator and we have introduced

\be
\beta^2=1-\frac{m_t^2}{s}\,\,\,.
\ee

\noindent We will consider the total cross section normalized to the
cross section for producing $\mu^+\mu^-$ pairs via one photon
exchange, i.e.

\be R^{tc} \equiv \frac{\sigma(e^+e^-\rightarrow t\bar c+ \bar
tc)}{\sigma( e^+e^- \rightarrow \gamma^*\rightarrow \mu^+\mu^-)}
\label{rtc_ee} \ee

\noindent and normalized to $\lambda_{ij}\simeq\lambda\!=\!1$ (see
Eq.~(\ref{coupl_sher})). For the moment, we want to simplify our
discussion by taking the same $\lambda$ for both $\xi^U_{tt}$ and
$\xi^U_{ct}$. Moreover, we want to factor out this parameter, because
it summarizes the degree of arbitrariness we have on these new
couplings and it will be useful for further discussion. We will
elaborate more about the possibility of considering different
alternatives for the FC couplings in Sec.~\ref{bsg}, after we
present a comprehensive analysis of the constraints.

\begin{figure}[htb]
\centering
\epsfxsize=5.in
\leavevmode\epsffile{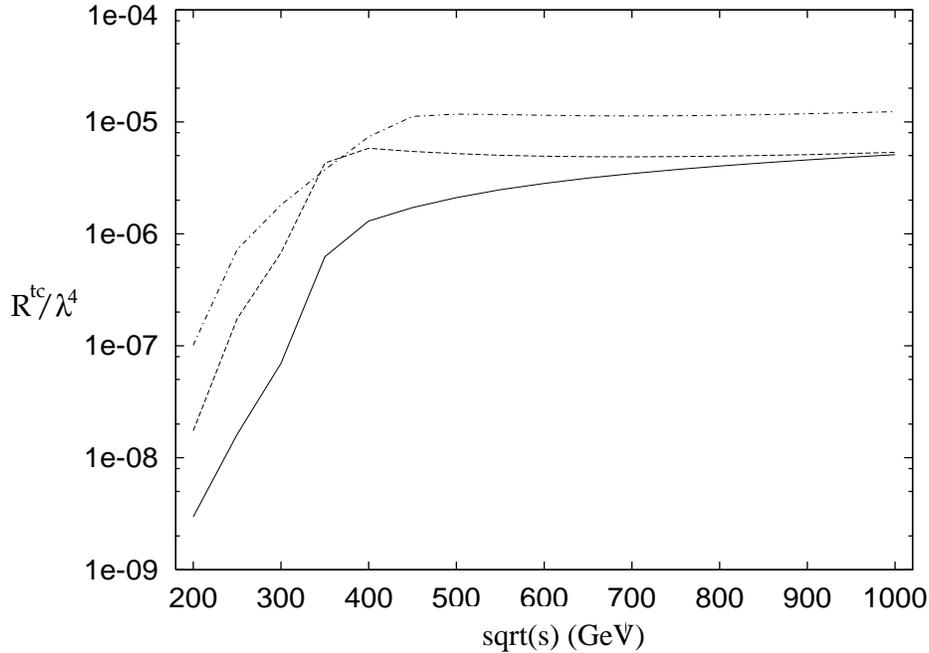}
\caption[]{ $R^{tc}/\lambda^4$ vs.\ $\sqrt{s}$ for case~1 (solid),
case~2 (dashed), and case~3 (dot-dashed), with $m_t=180$ GeV. See
Eq.~(\ref{threecases}).}
\label{eetc_plot}
\end{figure}

As already discussed in Ref.~\cite{eetc}, we take $m_t\!\simeq\!180$
GeV and vary the masses of the scalar and pseudoscalar fields in a
range between 200 GeV and 1 TeV\null. Larger values of the scalar
masses are excluded by the requirement of a weak-coupled scalar
sector. The phase $\alpha$ does not play a relevant role and, as we
discuss in Appendix~\ref{formfactors}, in our qualitative analysis we
will set $\alpha\!=\!0$. In this case, as we can read in
Eqs.~(\ref{masseigen}) or (\ref{nomasseigen}), $H^0\!=\!\bar H^0+v$,
$H^1\!=\!h^0$, and $H^2\!=\!A^0$ and the only new contributions come
from $h^0$ and $A^0$. In Fig.~\ref{eetc_plot} we plot
$R^{tc}/\lambda^4$ as a function of $\sqrt{s}$ for a sample of
relevant cases, in which one of the scalar particles is taken to be
light ($M_l\!\simeq\! 200$ GeV) compared to the other two
($M_h\!\simeq\!1$ TeV), i.e.

\bea
\label{threecases}
1)\,\,\,m_h &=& M_\ell \,\,\,\,\,\mbox{and}\,\,\,\,\, 
m_A\simeq m_c\simeq M_h\,; \nonumber\\
2)\,\,\,m_A &=& M_\ell\,\,\,\,\,\mbox{and}\,\,\,\,\, 
m_h\simeq m_c\simeq M_h\,;\\
3)\,\,\,m_c &=& M_\ell\,\,\,\,\,\mbox{and}\,\,\,\,\, 
m_h \simeq m_A\simeq M_h\,.\nonumber
\eea

\noindent where $m_h$ and $m_A$ are the neutral scalar and
pseudoscalar masses and $m_c$ is the charged scalar mass
respectively. We find that even with different choices of $m_h$, $m_A$
and $m_c$ it is difficult to push $R^{tc}/\lambda^4$ much higher than
$10^{-5}$. Therefore the three cases illustrated in
Fig.~\ref{eetc_plot} appear to be a good sample to illustrate the type
of predictions we can obtain for the rate for top-charm production in
model~III\null.

{}From Fig.~\ref{eetc_plot}, we also see that going to energies much
larger than $\sim 400$--500 GeV (i.e.\ $\sim2M_l$) does not gain much
in the rate and in this case $R^{tc}/\lambda^4$ can be as much as
$10^{-5}$. Since it is reasonable to expect $10^4$--$10^5$
$\mu^+\mu^-$ events in a year of running for the next generation of
$e^+e^-$ colliders ($\int{\cal L}\simeq 5\times
10^{33}\,\mbox{cm}^{-2}\mbox{sec}^{-1}$) at $\sqrt{s}=500$ GeV, this
signal could be at the detectable level only for not too small values
of the arbitrary parameter $\lambda$.  Thus we can expect experiments
to be able to constrain $\lambda\lsim1$, for scalar masses of a few
hundred GeVs.

\section{Rare top decays: $\lowercase{t}\rightarrow \lowercase{c}
\gamma,Z,\lowercase{g}$}
\label{tcgZ}

Starting from the form factors defined in Eq.~(\ref{oneloopvertex})
and given in Appendix~\ref{formfactors}, we can also easily derive the
rates for rare top decays like $t\rightarrow c\gamma$, $t\rightarrow
cZ$ and $t\rightarrow c g$. The study of rare top decays has been
often emphasized in the literature \cite{soni,gunion,hou,lukesavage},
in particular, as a potential source of evidence for new
physics. Indeed, as we can read from Table~\ref{raretop}, these decays
are extremely suppressed in the SM and they are quite small even in
the 2HDM's without tree level FCNC's (i.e.\ both in model~I and in
model~II) \cite{soni,gunion}. This is due to a strong GIM suppression
from the small value of the internal quark masses $m_{d,s,b}$ as well
as the large tree level rate for $t\rightarrow bW$. On the other hand,
these rare top decays normally get enhanced in models with FCNC's and
this motivates us to estimate their branching ratio in
model~III\null. From the experimental point of view the prospects for
the three modes, $t\rightarrow c\gamma$, $cZ$, and $cg$ are quite
different. In particular, $t\rightarrow cg$ could be quite problematic
for a hadron collider and the backgrounds will have to be considered
before one can ensure that they do not represent a serious
limitation. On the other hand, for the $e^+e^-$ case background issues
are less likely to be a serious problem even for $t\rightarrow cg$.

\begin{table}
\begin{center}
\begin{tabular}{c l c c c}
\\ Decay & SM & Model I & Model II & Model III \\ \\ \hline\\
$t\rightarrow c\gamma$ & $\sim 5\cdot 10^{-12}$ & 
$10^{-13}\!-\!10^{-11}$ & $10^{-13}\!-\!10^{-9}$ & 
$10^{-12}\!-\!10^{-7}$  \\ \\
$t\rightarrow cZ$ & $\sim 10^{-13}$ & $10^{-13}\!-\!10^{-11}$ &
$10^{-13}\!-\!10^{-10}$ & $10^{-8}\!-\!10^{-6}$ \\ \\
$t\rightarrow cg$ & $\sim 5\cdot 10^{-11}$ & $10^{-11}\!-\!10^{-9}$ &
$10^{-11}\!-\!10^{-8}$ & $10^{-8}\!-\!10^{-4}$ \\ \\
\end{tabular}
\caption[]{Values of $Br(t\rightarrow c\gamma)$, $Br(t\rightarrow
cZ)$, and $Br(t\rightarrow cg)$ for $m_t\simeq 180$ GeV, in the SM and
in the 2HDM's denoted as Model~I, Model~II, and Model~III\null. Each
range is obtained by varying $m_c$, $m_h$, $m_A$, $\tan\beta,\ldots$
over a broad region of the parameter space of the corresponding model,
as explained in the text. For Model~III, we have fixed
$\lambda_{ij}\simeq\lambda=1$ in the FC couplings.}
\label{raretop}
\end{center}
\end{table}

In model~III the modes $t\rightarrow c\gamma$ and $t\rightarrow cZ$
have been previously considered \cite{lukesavage} and their rates are
given by

\be
\Gamma(t\rightarrow c\gamma)=\frac{1}{(16\pi^2)^2}\frac{1}{8\pi}
\left(|C^\gamma|^2+|D^\gamma|^2\right) 
\ee

\bea
\Gamma(t\rightarrow c Z)&=&\frac{1}{(16\pi^2)^2}\frac{1}{16\pi
m_t}\left(1-\frac{M_Z^2}{m_t^2}\right)\left(\frac{m_t^2}{M_Z^2}-1\right)
\left[(m_t^2+2M_Z^2)\left(|A^Z|^2+|B^Z|^2\right)\right.\nonumber\\
&-&\left. 6M_Z^2\real(A^{Z*}C^Z-B^{Z*}D^Z)
+M_Z^2\left(\frac{M_Z^2}{m_t^2}+2\right)\left(|C^Z|^2+|D^Z|^2\right)
\right]\,\,\,.
\eea

\noindent The rate for $t\rightarrow cg$ can be written in an
analogous manner, as,

\be
\Gamma(t\rightarrow cg)=\frac{1}{(16\pi^2)^2}
\frac{1}{8\pi}C_F\left(|C^g|^2+|D^g|^2\right) 
\ee

\noindent where $C_F=(N^2-1)/2N$.

\begin{figure}[htb]
\centering
\epsfxsize=4.in
\leavevmode\epsffile{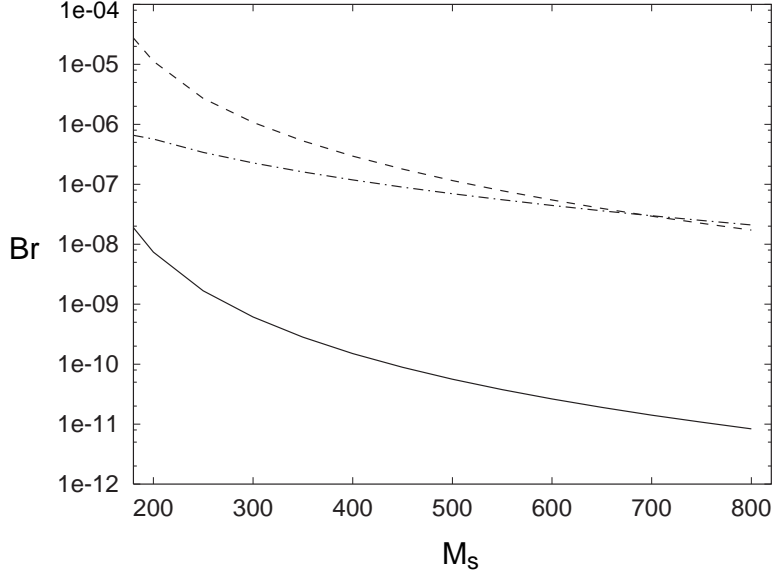}
\caption[]{ Branching fraction for $t\rightarrow c\gamma$ (solid),
$t\rightarrow cZ$ (dot-dashed), and $t\rightarrow cg$ (dashed) as a
function of a common scalar mass $M_s$, when $m_t=180$ GeV.}
\label{tc_phZg}
\end{figure}

\noindent The branching ratios reported in Table~\ref{raretop} are
obtained by normalizing $\Gamma(t\rightarrow c\gamma)$,
$\Gamma(t\rightarrow cZ)$ and $\Gamma(t\rightarrow cg)$, for
simplicity, just to the main decay $t\rightarrow bW$ rate, i.e.

\be
\Gamma(t\rightarrow bW)=\frac{G_F}{8\pi\sqrt{2}}|V_{tb}|^2m_t^3
\left(1-\frac{M_W^2}{m_t^2}\right)\left(1+\frac{M_W^2}{m_t^2}-
2\frac{M_W^4}{m_t^4}\right)\,\,\,.
\ee

\noindent The branching ratios for model~I and model~II are deduced
from the analysis of Ref.~\cite{soni,gunion}, with $m_t\!\simeq\!180$
GeV\null. The results for model~III are obtained by varying the
neutral and the charged scalar masses between 200 GeV and 1 TeV,
assuming different patterns as explained in the preceding section (see
for instance Eq.~(\ref{threecases})). In particular the upper bounds
on the different branching ratios given in Table~\ref{raretop} are
obtained by taking the scalars to have a common and relatively small
mass. We also notice that in model~III the results show a significant
dependence on $m_t$, such that the numbers in Table~\ref{raretop}
change on the average by as much as an order of magnitude when $m_t$
is varied between 150 GeV and 200 GeV\null. This sensitivity to $m_t$
may become relevant when the experiments ever get to the point of
being able to measure this type of rare top decay. Finally, in the FC
couplings of Eq.~(\ref{coupl_sher}) we also take all the
$\lambda_{ij}$ parameters to be equal to $\lambda$. In particular, the
numbers in Table~\ref{raretop} are given for $\lambda\!=\!1$.

Our analytical expressions for the form factors contain some
differences with respect to Ref.~\cite{lukesavage}, as explained in
Appendix \ref{formfactors} \cite{16pi}. Numerically they end up being
most relevant for $t\rightarrow c\gamma$. Fig.~\ref{tc_phZg}
illustrates the case in which a common value $M_s$ is taken for all
the scalar masses, as might be useful for comparison \cite{csi} with
Fig.~2 of Ref.~\cite{lukesavage}. We can see that the analytical
difference between us and Ref.~\cite{lukesavage} translates into a
numerical difference of more than one order of magnitude for the
$t\rightarrow c\gamma$ decay rate.

{}From Table~\ref{raretop}, we see that $Br(t\rightarrow c\gamma)$,
$Br(t\rightarrow cZ)$, and $Br(t\rightarrow cg)$ can be substantially
enhanced with respect both to the SM and to the 2HDM's with no FCSC's
(i.e.\ model~I and model~II). Depending on the size of the FC
couplings, in model~III we can gain even more than two of orders of
magnitude in each branching ratio. This is likely to make a crucial
difference at the next generation of lepton and hadron colliders where
a large number of top quarks will be produced. Therefore these
machines will be sensitive to signals from non-standard top decays and
should be able to put stringent bounds on the new interactions
involved, the FC couplings of model~III in our case. In view of these
future possibilities, a careful study of the FC couplings of the model
is mandatory and we will analyze the constraints that emerge from
existing experiments in Secs.~\ref{mixing}--\ref{bsg}.

\section{Top-charm production at $\mu^+\mu^-$ colliders}
\label{mumutc}

Another interesting possibility to study top-charm production is
offered by muon colliders \cite{mumutc}. Although very much in the
notion stage at present, a $\mu^+\mu^-$ collider has been suggested
\cite{cline,neuffer,barletta,palmer} as a possible lepton
collider. Muon colliders are especially interesting for two main
reasons. They can allow a detailed study of the $s$-channel Higgs and
also they may make it feasible to have high energy lepton colliders in
the multi-TeV regime. Neither of these goals is attainable with an
$e^+e^-$ collider.

If muon colliders are eventually shown to be a practical and desirable
tool, most of the applications would be very similar to electron
colliders. One additional advantage alluded to above, however, is that
they may be able to produce Higgs bosons ($\fH$) in the $s$-channel in
sufficient quantity to study their properties directly
\cite{cline,barger,bargertwo,mumu,mumutc}.  The crucial point is that
in spite of the fact that the $\mu^+\mu^-\fH$ coupling, being
proportional to $m_\mu$, is very small, if the muon collider is run on
the Higgs resonance, $\sqrt{s}=m_\fH$, Higgs bosons may be produced at
an appreciable rate \cite{cline,barger,bargertwo,mumu,mumutc}.

At $\sqrt{s}=m_\fH$, the cross section for producing $\fH$,
$\sigma_\fH$, normalized to $\sigma_0=\sigma(\mu^+\mu^-\rightarrow
\gamma^*\rightarrow e^+ e^-)$, is given by \cite{bargertwo,mumu}

\begin{equation}
R(\fH) = {\sigma_\fH \over \sigma_0} = {3\over\alpha_e^2} B_\mu^\fH
\label{rdef} 
\end{equation}

\noindent where $B_\mu^\fH$ is the branching ratio of $\fH\rightarrow
\mu^+\mu^-$ and $\alpha_e$ is the QED fine structure constant.  If
the Higgs is very narrow, the exact tuning to the resonance implied in
Eq.~(\ref{rdef}) may not in general be possible. The effective rate of
Higgs production will then be given by \cite{mumu}

\begin{equation}
\tilde R(\fH)=\left[ \frac{\Gamma_\fH}{m_\fH\delta} \arctan
\frac{m_\fH\delta}{\Gamma_\fH} \right] R(\fH) \label{rtildef}
\end{equation}

\noindent where it is  assumed that the energy of the beam has a
finite spread described by $\delta$

\begin{equation}
m_\fH^2(1-\delta)<s<m_\fH^2(1+\delta)
\end{equation}

\noindent and $s$ is uniform about this range. 

\begin{figure}[htb]
\centering
\epsfxsize=3.5in
\leavevmode\epsffile{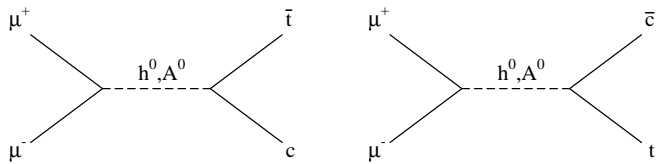}
\caption[]{Tree level contributions to $\mu^+\mu^-\rightarrow\bar
tc+t\bar c$ in model~III.}
\label{mumu_tree}
\end{figure}

In a recent paper \cite{mumutc} we have considered the simple but
fascinating possibility that such a Higgs, $\fH$, has a
flavor-changing $\fH t\bar c$ coupling, as is the case in model~III or
in any other 2HDM with FCNC's\null. The process $\mu^+\mu^-\rightarrow
t\bar c+\bar t c$ will then arise at the tree level as illustrated in
Fig.~\ref{mumu_tree}.  It will give a signal which should be easy to
identify, is likely to take place at an observable rate, and has a
negligible SM background. Thus the properties of the important $\fH
t\bar c$ coupling may be studied in detail.

For illustrative purposes we take $\fH=h^0$ in model~III where
$\alpha=0$ (case~1) or $\pi/4$ (case~2). The main distinction between
the two cases is that in case~2 the decays $h^0\rightarrow ZZ,\ WW$
are possible while in case~1 they are not (see Appendix~\ref{fr_mod3}
for the relevant Feynman rules). Thus case~1 is very similar to
$\fH=A^0$.  This will matter in computing the total width of the $h^0$
boson, i.e.\ the $Br(h^0\rightarrow tc)$, while it was completely
irrelevant in the $e^+e^-\rightarrow \gamma^*,Z^*\rightarrow t\bar
c+c\bar t$ calculation.

In general the FC coupling of $h^0$ to $t\bar c$ can be written as

\begin{equation}
C_{htc}= \frac{1}{\sqrt{2}}\left( \xi_{tc}P_R+\xi_{ct}^\dagger P_L \right)
\cos\alpha \equiv \frac{g\sqrt{m_t m_c}}{2 m_W}(\chi_R P_R+\chi_L P_L)
\label{htc}
\end{equation}

\noindent where $\chi_L$ and $\chi_R$ are in general complex numbers
and of order unity if Eq.~(\ref{coupl_sher}) applies. In particular we
will consider the case in which $\lambda_{ct}=\lambda\simeq O(1)$ (see
Eq.~(\ref{coupl_sher})) and $\chi_L$ and $\chi_R$ are real. We treated
the more general case in Ref.~\cite{mumutc} to which we refer for
further details.

The decay rate to $t\bar c$ is thus

\begin{equation}
\Gamma(\fH\rightarrow t\bar c)=  {3 g^2 m_t m_c m_\fH \over 
32\pi M_W^2} 
\left[ {(m_\fH^2-m_t^2)^2\over m_\fH^4} \right] \left(
{|\chi_R|^2+|\chi_L|^2\over 2} \right) \label{gamma_htc}
\end{equation}

\noindent and, $\Gamma(\fH\rightarrow t\bar c)=\Gamma(\fH\rightarrow
c\bar t)$ at the tree level that we are considering for now.

As we did in Sec.~\ref{eetc} for the $e^+e^-$ case, also in the
$\mu^+\mu^-$ case we can define the analogue of $R^{tc}$ in
Eq.~(\ref{rtc_ee}) to be

\begin{equation}
R_{tc}= \tilde R(\fH)\,(B^\fH_{t\bar c}+B^\fH_{c\bar t})
\label{rtc_mumu}
\end{equation}

\noindent where $B^\fH_{t\bar c}$ and $B^\fH_{c\bar t}$ are the
branching ratio for $\fH\rightarrow t\bar c$ and $\fH\rightarrow c
\bar t$ respectively. We estimate $R_{tc}$ in the following two cases:

\begin{enumerate}

\item [i)] case 1:  $\alpha=0$\,;

\item [ii)] case 2:  $\alpha=\pi/4\,\,.$

\end{enumerate}

\begin{figure}
\centering
\epsfxsize=4.in
\leavevmode\epsffile{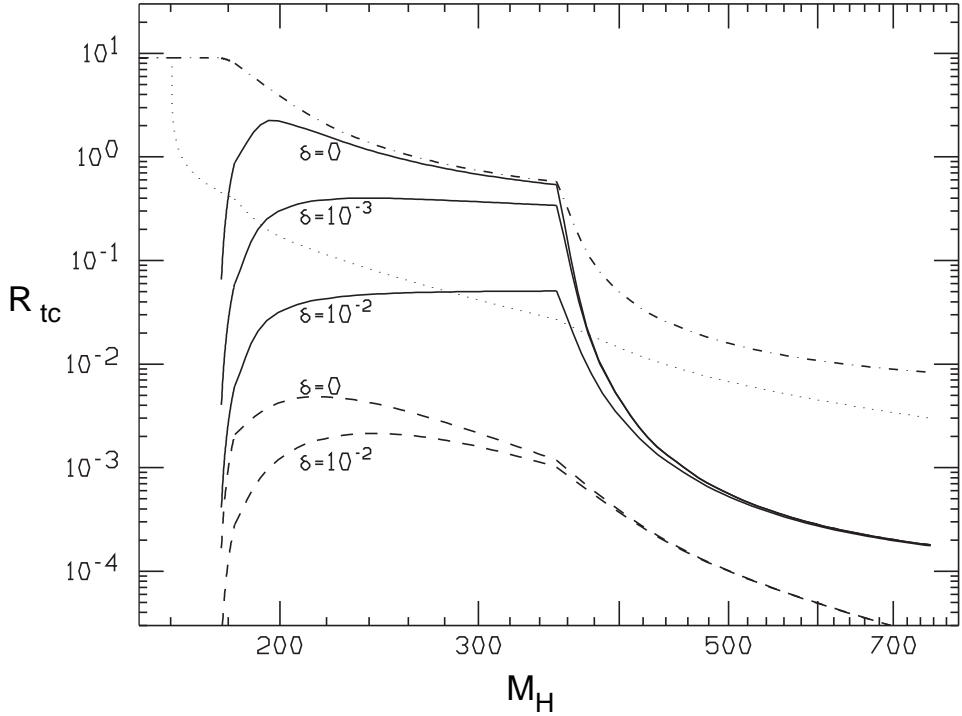}
\caption[]{$R_{tc}$ for $\delta=0$,
$10^{-3}$ and $10^{-2}$ in case~1 (set of solid curves) and case~2
(set of dashed curves). We also plot $\tilde R(\fH)$ in case~1
(dot-dashed) and case~2 (dotted).}
\label{mumu_all}
\end{figure}

\noindent Using the result in Eq.~(\ref{gamma_htc}) \cite{mumutc} and
taking the expressions for the standard partial widths for $\fH$
(e.g.\ $\Gamma(\fH\rightarrow t\bar t)$, $\Gamma(\fH\rightarrow b\bar
b), \ldots$) from the literature \cite{hunter}, we obtain the
following results.  In case~1, if $m_\fH$ is below the $t\bar t$
threshold, $R_{tc}$ is about $10^{-2}-1$ and in fact $tc$ makes up a
large branching ratio.  Above the $t\bar t$ threshold $R_{tc}$
drops. For case~2 the branching ratio is smaller due to the $WW$ and
$ZZ$ threshold at about the same mass as the $tc$ threshold and so
$R_{tc}$ is around $10^{-3}$. All these results are illustrated in
Fig.~\ref{mumu_all}, where we plot $\tilde R(\fH)$ and $R_{tc}$ with
$\delta=0$, $10^{-3}$ and $10^{-2}$ in case~1 and case~2~.

For a specific example, let us take $m_\fH=300$ GeV,
i.e. $\sigma_0\approx 1$ pb. For a luminosity of $10^{34}
\mbox{cm}^{-2} \mbox{s}^{-1}$, a year of $10^7$ s ($1/3$ efficiency)
and for $\delta=10^{-2}$, case~1 will produce about $5\times10^3
(t\bar c+\bar t c)$ events and case~2 will produce about $150$
events. Given the distinctive nature of the final state and the lack
of a SM background, sufficient luminosity should allow the observation
of such events.

If such events are observed, the $\mu^+\mu^-$ collider offers the
additional interesting possibility of extracting the values of the
$\chi_L$ and $\chi_R$ couplings in Eq.~(\ref{htc}) separately. What is
measured initially at a $\mu^+\mu^-$ collider is $R_{tc}$. One
is required to know the total width of the $\fH$ and the energy spread
of the beam in order to translate this into $\Gamma(\fH\rightarrow t\bar
c)$. This then allows the determination of $|\chi_L|^2+|\chi_R|^2$
(see Eq.~(\ref{gamma_htc})). To get information separately on the two
couplings we note that the total helicity of the top quark is

\begin{equation}
{\bf H}_t=-{\bf H}_{\bar t}= {|\chi_R|^2-|\chi_L|^2  
\over |\chi_R|^2+|\chi_L|^2}
\end{equation}

\noindent from which one may therefore infer $|\chi_L|$ and
$|\chi_R|$. Of course the helicity of the $t$ cannot be observed
directly. However, following the discussion of \cite{mumu,mumutc} one
may obtain it from the decay distributions of the top quark.
Unfortunately in the limit of small $m_c$ the helicity of the
$c$-quark is conserved. Hence the relative phase of $\chi_L$ and
$\chi_R$ may not be determined since the two couplings do not
interfere.

\section{Constraints from $F^0\!\!-\!\!\bar F^0$ mixing processes}
\label{mixing}

{}From the previous analysis we see that both
$e^+e^-\,,\,\mu^+\mu^-\rightarrow\bar c t+\bar t c$ and $t\rightarrow
c\gamma,Z,g$ could be of some experimental relevance depending on the
size of the FC couplings of model~III\null.  As is well known from the
literature on FCNC's, the most dangerous constraints on tree level FC
couplings come usually from $F^0\!-\!\bar F^0$ mixing processes
($F=K,B_d,D$) \cite{sher,antaramian,hall}. In these references we can
find the bounds imposed on some tree level FC couplings by different
$F^0\!-\!\bar F^0$ mixing processes. Due to the specific structure of
the couplings of model~III and to the new phenomenology at hand, we
think that a more careful analysis is due, which takes into account
both tree level and loop contributions. We have examined the $\Delta
F\!=\!2$ mixing processes in detail and concluded that both the
$K^0\!-\!\bar K^0$ and the $B^0_d\!-\!\bar B^0_d$ mixings are
particularly effective in constraining some FC couplings, while the
experimental determination of the $D^0\!-\!\bar D^0$ mixing is, for
now, not good enough to compete with the other two mixings. However,
due to the different flavor structure of the $D^0\!-\!\bar D^0$
mixing, it would be extremely important to have a good experimental
determination in this case as well. We will address the problem more
specifically later on in this section.

In a model with FCNC's, $F^0\!-\!\bar F^0$ mixing processes can arise
at the tree level. Therefore they are likely to be greatly enhanced
with respect to the SM where they appear only at the one loop
level. Due to the good agreement between the SM prediction and the
experimental determination of the mass difference in the $K^0\!-\!\bar
K^0$ and $B^0_d\!-\!\bar B^0_d$ systems, any tree level contribution
from elementary FC couplings needs to be strongly suppressed. This was
the original motivation for imposing on the 2HDM's a discrete symmetry
\cite{glash} which could prevent tree level FCNC's from appearing
(model~I and model~II). Our goal will be now to verify if, in
model~III, the hierarchy imposed on the $\xi_{ij}^{U,D}$ couplings by
the ansatz in Eq.~(\ref{coupl_sher}) is strong enough to make the tree
level contribution to any $F^0\!-\!\bar F^0$ mixing sufficiently small
to be compatible with the experimental constraints.

\begin{figure}
\centering
\epsfxsize=4.5in
\leavevmode\epsffile{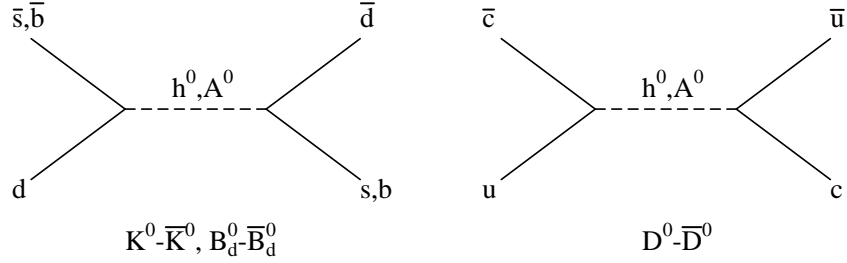}
\caption[]{Tree level contributions to each different $F^0\!-\!\bar
F^0$ mixing, for $F=K,B_d,D$, in model~III.}
\label{treelevel}
\end{figure}

\begin{figure}
\centering
\epsfxsize=5.5in
\leavevmode\epsffile{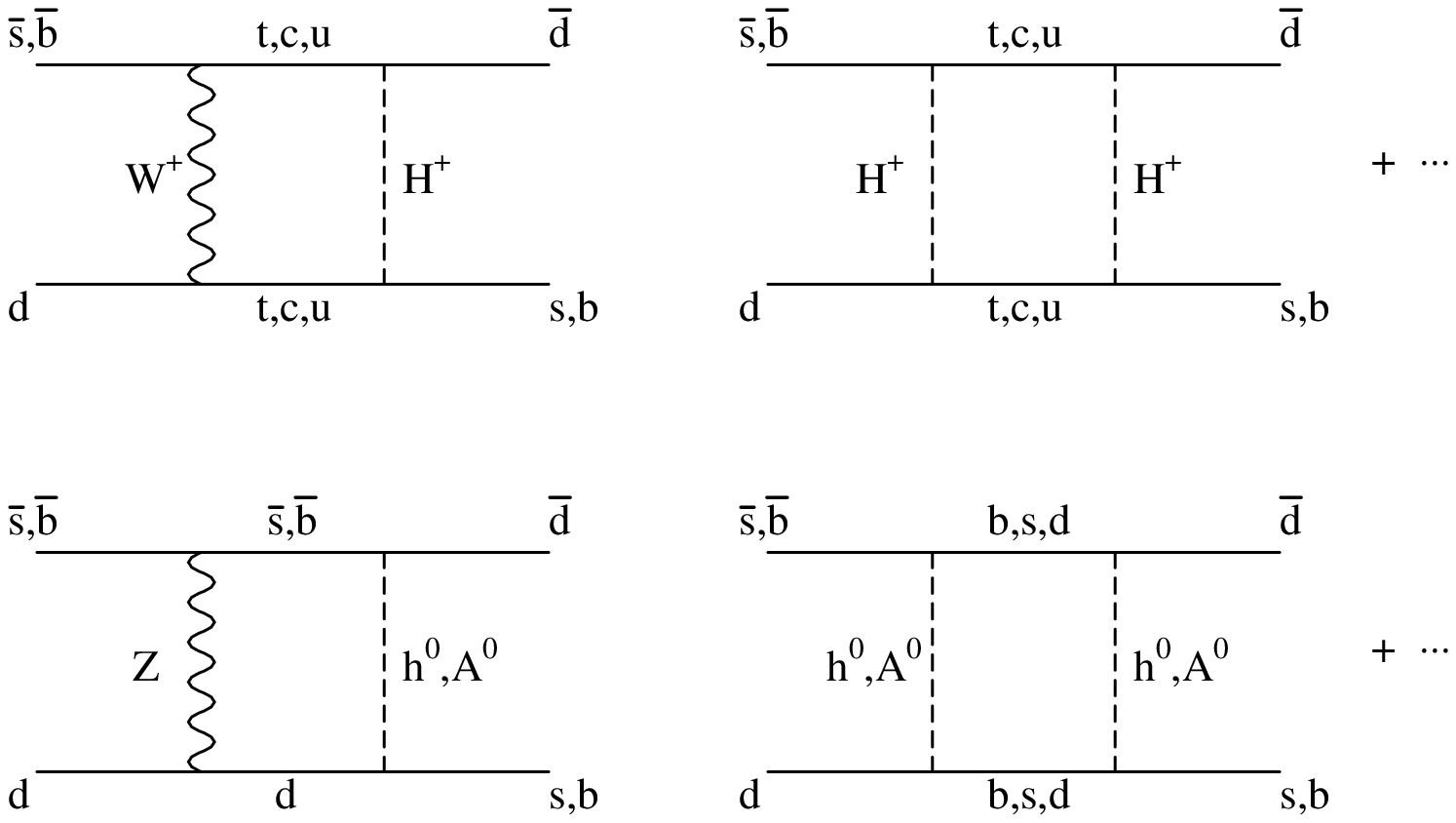}
\caption[]{Box diagrams which contribute at one loop to the
$K^0\!-\!\bar K^0$ and to the $B^0_d\!-\!\bar B_d^0$ mixing, in
model~III\null. The $D^0\!-\!\bar D^0$ case is obtained by
appropriately replacing the external and internal quark states.}
\label{1loop_box}
\end{figure}

\begin{figure}[htb]
\centering
\epsfxsize=5.in
\leavevmode\epsffile{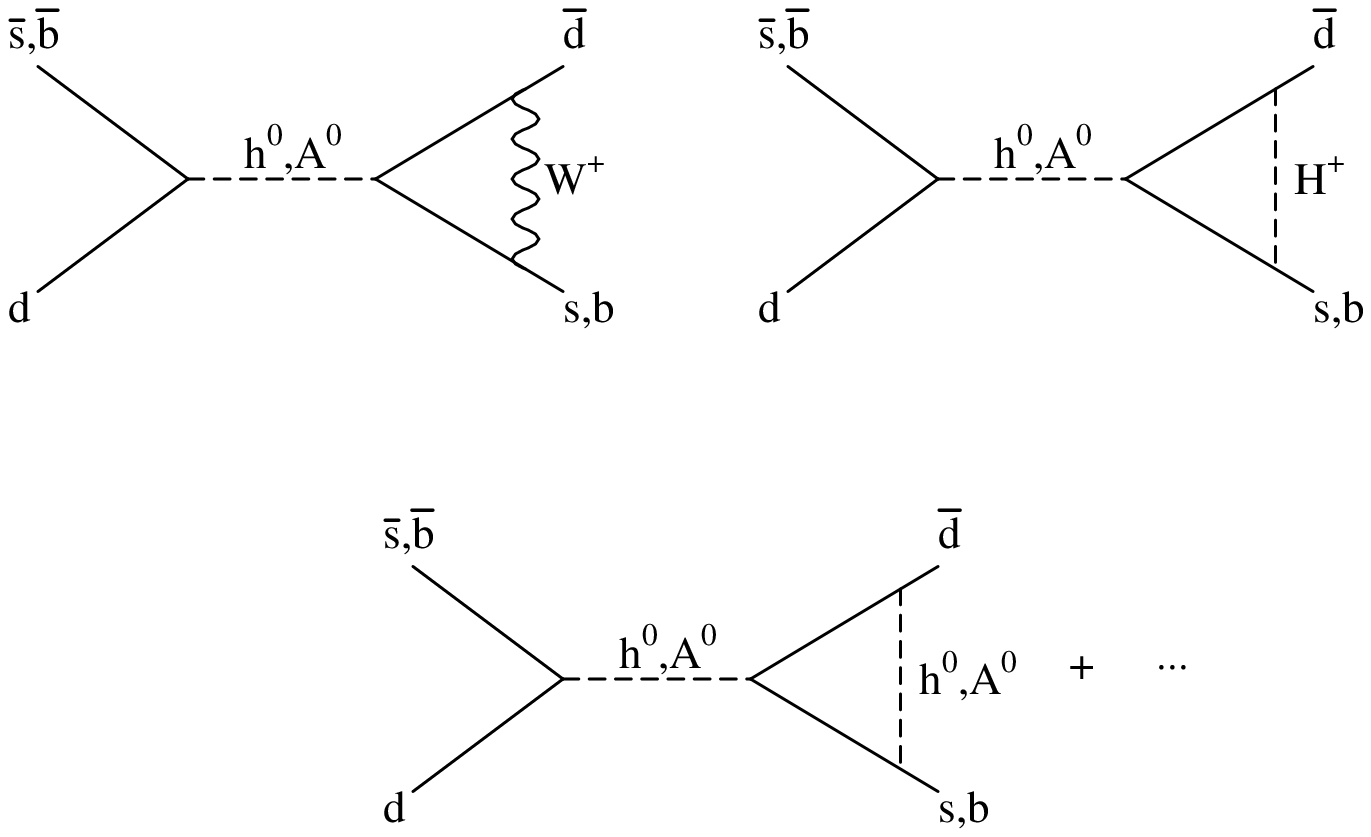}
\caption[]{Penguin diagrams which contribute at one loop to the 
$K^0\!-\!\bar K^0$  and to the $B^0_d\!-\!\bar B_d^0$ mixing, in 
model~III\null. The $D^0\!-\!\bar D^0$ case is obtained by appropriately
replacing the external and internal quark states.}
\label{1loop_penguin}
\end{figure}

For any $F^0\!-\!\bar F^0$ mixing we have evaluated, both at the tree
level and at the one loop level, the mass difference between the mass
eigenstates of the system, given respectively by

\bea
M_K \Delta M_K&\simeq&\mbox{Re}\langle K^0|(\bar s d)_{V-A}(\bar s
d)_{V-A}|\bar K^0\rangle\nonumber\\
M_{B_d} \Delta M_{B_d}&\simeq&|\langle B^0_d|(\bar b d)_{V-A}(\bar b
d)_{V-A}|\bar B^0_d\rangle|\\
\label{dm_th}
M_D \Delta M_D&\simeq&|\langle D^0|(\bar c u)_{V-A}(\bar c
u)_{V-A}|\bar D^0\rangle|\nonumber 
\eea

\noindent where we use the notation $(\bar q q^\prime)_{V-A}=\bar
q\gamma^{\mu}(1-\gamma^5)q^\prime$. The tree level contributions for
each different mixing are shown in Fig.~\ref{treelevel}, while a
sample of one loop contributions are illustrated in
Figs.~\ref{1loop_box} and \ref{1loop_penguin}, for the box and the
penguin diagrams respectively.

In our analysis we have made some general approximations which we want
to discuss first.

\begin{enumerate}

\item [i)] We observe that, as in the calculation of the $Ztc$ and
$\gamma tc$ form factors in Sec.~\ref{eetc}, also in this case the
value of the phase $\alpha$ plays a minor role numerically.
Therefore, as we did in Sec.~\ref{eetc}, we will set $\alpha\!=\!0$ in
the following analysis.  Once that is done we can focus our attention
only on the contributions coming from $H^1\!=\!h^0$, $H^2\!=\!A^0$,
and $H^+$. The possibility of having FCNC's induced by $h^0$ and $A^0$
will give rise to the tree level $\Delta F\!=\!2$ mixings of
Fig.~\ref{treelevel}.

\item [ii)]In the evaluation of the one loop contributions to the
$\Delta F\!=\!2$ mixings, we will take advantage of the fact that they
are due to scalar bosons whose couplings to the quark fields are
proportional to the masses of the quarks involved and, for the charged
scalar, to some combinations of CKM matrix elements. Therefore, in
each case we will consider only the dominant contribution, which quite
often will correspond to the diagrams with a heavy quark loop. This
procedure is clearly more approximate than the exact calculation one
uses to perform in the SM \cite{inamilim}, but a similar order of
accuracy would not be necessary in our case. Nevertheless, as a check
of our calculation, we have reproduced the SM result for each mixing
and used it as a reference point to fix the right relative signs and
normalizations.

\item [iii)]The $\Delta F\!=\!2$ effective interactions generated by
the new scalar fields are often more complicated compared to the SM
results \cite{inamilim}, because the scalar-fermion couplings involve
more chiral structures. Therefore, the evaluation of the mass
difference in the various $F^0\!-\!\bar F^0$ systems will involve the
matrix elements of operators other than just the SM one

\be
O^{SM}_{\Delta F\!=\!2}=O^F_{VLL}=(\bar f\gamma^{\mu}(1-\gamma_5)q) 
(\bar f\gamma_{\mu}(1-\gamma_5)q)\,\,.
\ee 

\noindent In general, matrix elements of the following operators
are involved

\bea
O_S^F&=&(\bar f q)(\bar f q)\nonumber\\
O_P^F&=&(\bar f\gamma_5 q)(\bar f\gamma_5 q)\nonumber\\
O_V^F&=&(\bar f\gamma^{\mu}q)(\bar f\gamma_{\mu}q)\nonumber\\
O_A^F&=&(\bar f\gamma^{\mu}\gamma_5 q)(\bar f\gamma_{\mu}\gamma_5q)\\
\label{newop}
O_{VLR}^F&=&(\bar f\gamma^{\mu}(1-\gamma_5)q)
(\bar f\gamma_{\mu}(1+\gamma_5)q)\nonumber\\
O_{LL}^F&=&(\bar f(1-\gamma_5)q)(\bar f(1-\gamma_5)q)\nonumber\\
O_{LR}^F&=&(\bar f(1-\gamma_5)q)(\bar f(1+\gamma_5)q)\nonumber
\eea

\noindent for $F=K,B_d,D$; $f=s,b,c$ and $q=d,u$ depending on the
$F^0$ meson that we consider. The matrix element of $O^{SM}_{\Delta
F\!=\!2}$ is usually given as

\be 
\langle F^0|O^{SM}_{\Delta F\!=\!2}|\bar F^0\rangle= 
B_F\langle F^0|O^{SM}_{\Delta F\!=\!2}|\bar F^0\rangle_{VIA} 
\ee

\noindent where the ratio of the matrix element itself to its value in
the Vacuum Insertion Approximation (VIA) is expressed by the
$B$-parameter $B_F$. Extensive non-perturbative studies of $B_K$ and
$B_B$ exist in the literature, especially from lattice calculations
\cite{bpar}. For our purpose, we just want to evaluate each matrix
element in the VIA and use a common $B$-parameter (the one for
$O^F_{VLL}$). Clearly this is only an approximation. Such an
approximation may be problematic in the SM, where one aims to get a
very precise prediction, but it suffices for our qualitative
discussion. In particular, we will assume $B_K\!=\!0.75$, $B_B\!=\!1$,
and $B_D\!=\!1$ \cite{bpar}. Moreover, according to
Ref.~\cite{shanker} and to analogous calculations we performed for
those operators that were not considered there, we will use the
following expressions for the $\Delta F\!=\!2$ matrix elements of the
$O_a^F$ operators (for $a=S,P,V,A,VLL,VLR,LL,LR$) in the VIA

\bea 
M_S^F&=&\langle F^0|O^F_S|\bar F^0\rangle_{\sss VIA}=
-\frac{1}{6}M^{0,F}_P+\frac{1}{6}M^{0,F}_A \nonumber\\ 
M_P^F&=&\langle F^0|O^F_P|\bar F^0\rangle_{\sss VIA}=
\frac{11}{6}M^{0,F}_P-\frac{1}{6}M^{0,F}_A \nonumber\\
M_V^F&=&\langle F^0|O^F_V|\bar F^0\rangle_{\sss VIA}=
\frac{2}{3}M^{0,F}_P+\frac{1}{3}M^{0,F}_A \nonumber \\
M_A^F&=&\langle F^0|O^F_A|\bar F^0\rangle_{\sss VIA}=
-\frac{2}{3}M^{0,F}_P+\frac{7}{3}M^{0,F}_A \\ 
\label{matrix elements}
M_{VLL}^F&=&\langle F^0|O^F_{VLL}|\bar F^0\rangle_{\sss VIA}=
\frac{8}{3}M^{0,F}_A \nonumber \\ 
M_{VLR}^F&=&\langle F^0|O^F_{VLR}|\bar F^0\rangle_{\sss VIA}=
\frac{4}{3}M^{0,F}_P-2M^{0,F}_A \nonumber \\ 
M_{LL}^F&=&\langle F^0|O^F_{LL}|\bar F^0\rangle_{\sss VIA}=
\frac{5}{3}M^{0,F}_P \nonumber \\
M_{LR}^F&=&\langle F^0|O^F_{LR}|\bar F^0\rangle_{\sss VIA}=
-2M^{0,F}_P+\frac{1}{3}M^{0,F}_A \nonumber
\eea

\noindent for $F=K,B_d,D$. All the previous matrix elements have been
expressed in terms of the matrix elements of the only two operators which
do not vanish on the vacuum, i.e.

\bea
M^{0,F}_P&=& \langle\bar F^0|\bar\psi_f\gamma_5\psi_q|0\rangle
\langle 0|\bar\psi_f\gamma_5\psi_q|F^0\rangle = 
-f_F^2 \frac{M_F^4}{(m_f+m_q)^2}\\
M^{0,F}_A &=& \langle\bar F^0|\bar\psi_f\gamma_{\mu}\gamma_5\psi_q
|0\rangle\langle 0|\bar\psi_f\gamma^{\mu}\gamma_5\psi_q|F^0\rangle = 
f_F^2 M_F^2 \nonumber 
\eea

\noindent where $M_F$ and $m_f$ indicate respectively the mass of the
meson and of the quark of flavor $f$.  The recent evaluation of the
pseudoscalar decay constants $(f_F)$ can be found in the literature
\cite{ff}. In our calculation we have used the following set of
values: $f_K=0.160$ GeV, $f_B=0.175$ GeV, and $f_D=0.200$ GeV, where
the first value comes from experiments and the last two are
representatives of lattice calculations.

\item [iv)] No QCD corrections have been taken into account for
model~III\null. From the SM case \cite{buras_bb,nierste} we know that
these corrections can be important when a precise comparison with the
experimental data is needed \cite{qcd}. However, in model~III they
would affect the evaluation of the constraints at a much higher degree
of accuracy compared to the approximations that we are
adopting; therefore we neglect them.

\end{enumerate}

We now proceed to the discussion of the important results.  In
Eq.~(\ref{coupl_sher}) we basically parametrize our ignorance of the
FC couplings of model~III introducing the $\lambda_{ij}^{U,D}$ mixing
parameters.  Therefore, the constraints we are going to impose will
allow us to deduce the order of magnitude of some of the
$\lambda_{ij}^{U,D}$. Due to the fact that the analysis of any
$F^0\!-\!\bar F^0$ mixing will involve both tree level and one loop
contributions, we will have to deal with several couplings at the same
time. In order to simplify the analysis we will first take all the
$\lambda_{ij}^{U,D}$ parameters to be equal. Depending on the result
of this first approach to the problem, we will consider the
possibility that different couplings are differently enhanced or
suppressed. According to this logic, we have considered the following
cases, corresponding to three possible assumptions on the FC couplings
of Eq.~(\ref{coupl_sher}).

\begin{description}

\item [{\bf Case 1}] : $\lambda_{ij}\simeq\lambda$ common to all the FC
couplings.

\item [{\bf Case 2}] : $\lambda_{ui},\lambda_{dj}\ll 1$ for $i,j=1,2,3$,
i.e.\ negligible FC couplings for the first generation and no
assumptions on the other FC couplings.

\item [{\bf Case 3}] : as Case~2 but with the further assumption that

\be
\lambda_{bb},\lambda_{sb}\gg 1 \,\,\,\,\mbox{and}\,\,\,\,
\lambda_{tt},\lambda_{ct}\ll 1\,\,.
\nonumber
\ee

\end{description}

\noindent The results for each $F^0\!-\!\bar F^0$ mixing ($F=K,B_d,D$)
in case~1, case~2, and case~3 are reported in Table~\ref{f0f0}, where
we also give the corresponding experimental results \cite{eps,pdg} and
the SM predictions \cite{nierste,buras_bb,ohl}. Both for the SM and
for each case of model~III we also specify whether the dominant
contribution is due to tree level or to one loop diagrams. The
different relevance of the tree level and one loop contributions and
how this imposes constraints on some particular couplings will be
explained as we go along the discussion in this section. Moreover, for
each different choice of the couplings, we have varied $m_h$, $m_A$,
and $m_c$ in the range between 100 GeV and 1 TeV\null. This is the
reason for the range of values that are given in Table~\ref{f0f0} for
each case and for each mixing in model~III\null. In this section we
will discuss in particular case~1 and case~2. The scenario described
by case~3 results from the constraints imposed by $R_b$, $\rho$ and
$Br(B\rightarrow X_s\gamma)$ and will be discussed in some detail in
Sec.~\ref{mixing}.  For convenience, a summary of its results is
reported in Table \ref{f0f0} as well.

In {\bf Case 1}, i.e.\ when all the FC couplings are parametrized in
terms of a unique $\lambda$, the leading contribution comes
from the tree level diagrams of Fig.~\ref{treelevel}. The one loop
contributions are always subleading all over the mass parameter
space. For each $F^0\!-\!\bar F^0$ mixing, there are two possible tree
level contributions, mediated by an $h^0$ and an $A^0$ neutral field
respectively. We want to consider them separately because in our
analysis we will vary $m_h$ and $m_A$ independently. Moreover the two
contributions differ by the chiral structure of the resulting four
fermion effective interaction, of $S$-type for the $h^0$ exchange and
of $P$-type for the $A^0$ exchange (see their Feynman rules in
Appendix \ref{fr_mod3}). Since $M^F_P\gg M^F_S$, the distinction
between the two tree level contributions becomes important and we
write them as follows

\begin{table}
\begin{center}
\begin{tabular}{c c c c c}
\\ & & 
$K^0\!-\!\bar K^0$ & 
$B^0_d\!-\!\bar B^0_d$ & 
$D^0\!-\!\bar D^0$ \\ \\ \hline\\
\multicolumn{2}{c}{Experiment} & $3.51\cdot 10^{-15}$ & 
$3.26\cdot 10^{-13}$ & $<1.32\cdot 10^{-13}$ \\ \\
\parbox{2cm}{\centering SM} & \parbox{1.3cm}{\centering One Loop} & 
$1.4-4.6\cdot 10^{-15}$ & $10^{-13}-10^{-12}$ & $10^{-17}-10^{-16}$ \\ \\
\parbox{2cm}{\centering Model III (Case 1)} & 
\parbox{1.3cm}{\centering Tree Level} & 
$10^{-14}-10^{-13}$ & $10^{-12}-10^{-11}$ & $10^{-13}-10^{-12}$ \\ \\
\parbox{2cm}{ \centering Model III (Case 2)} & 
\parbox{1.3cm}{\centering One Loop} & 
$10^{-18}-10^{-17}$ & $10^{-14}-10^{-13}$ & $10^{-17}-10^{-18}$ \\ \\
\parbox{2cm}{ \centering Model III (Case 3)} & 
\parbox{1.3cm}{\centering One Loop} & 
$10^{-20}-10^{-19}$ & $10^{-17}-10^{-16}$ & $10^{-14}-10^{-15}$ \\ \\
\end{tabular}
\caption[]{Experimental values and theoretical predictions in the SM
and in model~III for $\Delta M_F$ (in GeV) for different $F^0-\bar
F^0$ mixings ($F=K,B_d,D$). Case~1, case~2, and case~3 correspond to
three possible scenarios in which different assumptions are made on
the FC couplings, as described in the text. Each range is obtained by
varying the parameters of the model over a large region of the
parameter space, compatible with phenomenology and with the assumption
of weakly coupled scalar fields. The leading contributions (tree level
or one loop) are also indicated in each case.}
\label{f0f0}
\end{center}
\end{table}

\bea
\label{f0f0_tree}
h^0 &\rightarrow &\Delta M_F^{\sss tree}=
2\left(\xi^{U,D}_{fq}\right)^2\frac{1}{m_h^2}M_S^F\\ 
A^0 &\rightarrow & \Delta M_F^{\sss tree}=-2\left(\xi^{U,D}_{fq}
\right)^2\frac{1}{m_A^2}M_P^F\nonumber 
\eea

\noindent for $F=K,B,D$, $f=s,b,c$, and $q=d$ or $u$ depending on the
mixing that is of interest. In particular, since $M^F_P\gg M^F_S$, we
observe that an interesting possibility will be to have a light $h^0$
and a much heavier $A^0$. In this way the $A^0$ contribution would not
be too large, while we would still have a light scalar field with FC
interactions.

\begin{figure}[htb]
\centering
\epsfxsize=4.in
\leavevmode\epsffile{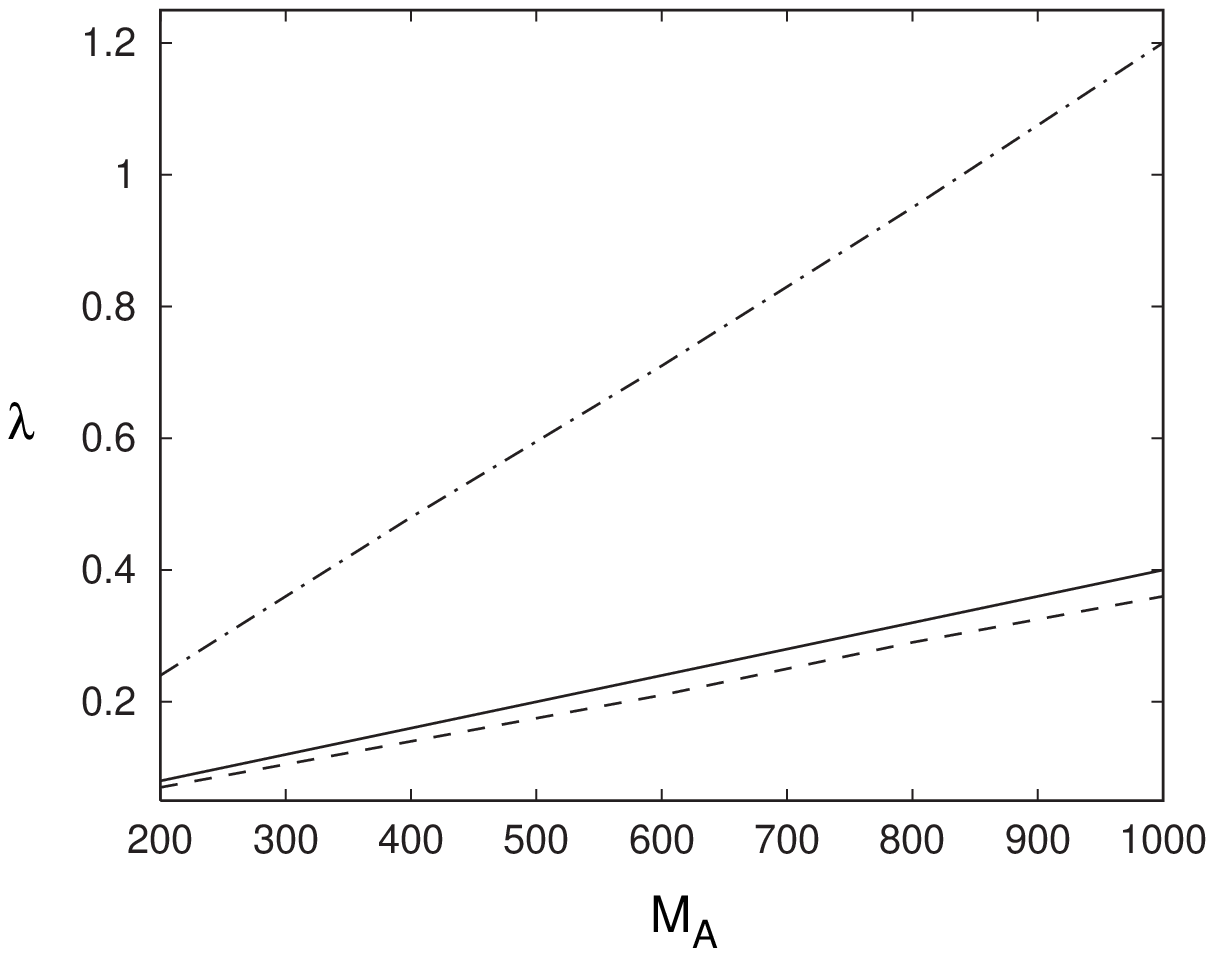}
\caption[]{Upper bounds on $\lambda$ imposed by the tree level mixings
illustrated in Fig.~\ref{treelevel}: $K^0\!-\!\bar K^0$ (dashed),
$B^0_d\!-\!\bar B^0_d$ (solid), and $D^0\!-\!\bar D^0$ (dot-dashed).}
\label{lambda1}
\end{figure}

In Fig.~\ref{lambda1} we illustrate the constraints imposed on the
parameter $\lambda$ by the tree level mixings in the $K^0\!-\!\bar
K^0$, $B^0_d\!-\!\bar B^0_d$, and $D^0\!-\!\bar D^0$ case respectively.
We recall that $\lambda$ is common to the three FC couplings which
govern the tree level mixings

\be
\xi^D_{sd}=\lambda\frac{\sqrt{m_sm_d}}{v}\,\,\,,\,\,\,
\xi^D_{bd}=\lambda\frac{\sqrt{m_bm_d}}{v}\,\,\,,\,\,\,
\xi^U_{cu}=\lambda\frac{\sqrt{m_cm_u}}{v}
\ee

\noindent The three curves represent the upper bounds on $\lambda$
imposed by the present experimental results (see Table \ref{f0f0}) for
different values of the neutral pseudoscalar mass $m_A$, when we fix
$m_h=200$ GeV\null. Lighter values of $m_h$ would not give any
relevant change to these bounds. Therefore, for each value of $m_A$,
the region above a given curve is ruled out by the corresponding
$F^0\!-\!\bar F^0$ mixing.  As we can see, the most relevant role is
played by $K^0\!-\!\bar K^0$ and $B^0_d\!-\!\bar B^0_d$, which
constrain $\lambda$ to be definitely smaller than unity, even for
large values of $m_A$ (i.e.\ $m_A\sim$ 1 TeV). We have verified that
if the experimental precision on $D^0\!-\!\bar D^0$ were increased by
one order of magnitude, this mixing would also start to play
approximately the same role as $K^0\!-\!\bar K^0$ and $B^0_d\!-\!\bar
B^0_d$, so that the three lines in Fig. \ref{lambda1} would then
roughly collapse into one line. We thus see that the $\Delta F\!=\!2$
mixings put severe bounds on the magnitude of the $\xi^{U,D}_{ij}$
couplings when we require all of them to be proportional to a common
$\lambda$. For $\lambda\le 0.1$ (as in Fig.~\ref{lambda1} for
$m_A\sim200$ GeV) it becomes difficult for the FC couplings of
model~III to play a role in any process. For instance, the case of
top-charm production that we discussed in Sec.~\ref{eetc} as well as
any rare top decay would be too suppressed to be of experimental
relevance.  On the other hand there is no good reason to assume that
all the $\lambda_{ij}$ of Eq.~(\ref{coupl_sher}) are equal.

Therefore, let us consider next the more general case in which the
$\lambda_{ij}$ parameters are different for each different
coupling. This is indeed the scenario that we identified as {\bf
Case~2}. We do not have enough specific measurements to constrain all
of them independently. However we can make some general remarks. We
first observe that the tree level mixings constrain only three
couplings to be necessarily small: $\xi_{sd}^D$, $\xi_{bd}^D$ and
$\xi_{uc}^U$.  In particular, we can take the point of view that the
bounds on $\lambda$ in Fig.~\ref{lambda1} do apply only to
$\lambda_{sd}$, $\lambda_{bd}$, and $\lambda_{uc}$. In general, we can
assume that all the FC couplings involving the first family, including
also $\xi^U_{ut}$, are suppressed.  In case~2, the tree level mixings
are suppressed enough that some loop contributions might become
important.  We will consider a loop contribution to become important
when it is clearly bigger than the corresponding SM prediction,
because only in this case it becomes possible to deduce a strong bound
on the new FC couplings.

The one loop diagrams that are most likely to be relevant are those
which involve charged scalars only, because the charged couplings can
also contain terms that do not involve any of the suppressed couplings
(see Eq.~(\ref{charged})) of the first-family. This is not the case
for the penguin-like diagrams of Fig.~\ref{1loop_penguin}, which
indeed turn out to be negligible, except for very light and
unrealistic values of the neutral scalar and pseudoscalar masses. We
mainly have to focus on the box diagrams of Fig.~\ref{1loop_box} with
charged scalars. Moreover, we notice that, if of comparable size, the
most relevant box diagrams are going to be those with only one (rather
than two) charged scalar boson because they are proportional to the
product of two $\lambda_{ij}$ instead of four $\lambda_{ij}$. Thus, if
quantitatively relevant with respect to the corresponding SM
contributions, they may be as effective as the tree level mixing in
constraining the $\lambda$'s.

For the $B^0_d\!-\!\bar B^0_d$ mixing, the only sizable contribution
comes from the diagrams with a top quark in the loop, because the
scalar field couples to the quarks as given by Eq.~(\ref{coupl_sher}),
i.e.\ proportionally to their masses.  For $K^0\!-\!\bar K^0$ mixing,
the diagram with a charm quark in the loop can be even more important
than the one with a top quark, because of the CKM couplings involved.
Finally, in the $D^0\!-\!\bar D^0$ case, the most relevant loop
contributions come from the box diagram with a bottom quark in the
loop. For very light $m_h$ and $m_A$, the penguin diagrams generated
by $h^0$ and $A^0$, with a top loop, can be comparable.

The results of this analysis can be read off Table~\ref{f0f0}.  In
spite of the high degree of arbitrariness that we have in model~III,
we do not find any region of the parameter space in which the one loop
contributions in case~2 are definitely larger than the SM
prediction. In other words, once the first family is assumed to
decouple, the $\Delta F\!=\!2$ processes for $K^0\!-\!\bar K^0$,
$B_d^0\!-\!\bar B_d^0$ and $D^0\!-\!\bar D^0$ place no further
constraints on the remaining FC couplings (i.e. $\xi^U_{ct}$ and
$\xi^D_{sb}$). Therefore this analysis tells us that case~2 is
compatible with the existing experimental measurements of the $\Delta
F\!=\!2$ mixings as long as the second and third generation FC
couplings of model~III do not have a $\lambda_{ij}$ much larger than
1. This is an interesting result as far as our predictions for
$\sigma(e^+e^-\rightarrow \bar t c +\bar c t)$ are concerned. Due to
the constraints we cannot enhance this cross section in any
substantial fashion, but at least it is not suppressed and the values
shown in Fig.~\ref{eetc_plot} are roughly correct, with some room for
a small enhancement. Similar considerations apply to the case of the
rare top decays that we discussed in Sec.~\ref{tcgZ}.

Other more complicated assumptions on the FC couplings of the second
and third generation could also be investigated. In order to consider
only those possibilities that respect the best fit of the available
phenomenological constraints, we will postpone the analysis of other
scenarios after the discussion of additional constraints from
$Br(B\rightarrow X_s\gamma)$, the $\rho$ parameter, and $R_b$, in
Sec.~\ref{bsg}.  As we have anticipated in presenting the results of
Table~\ref{f0f0}, we will discuss in particular another scenario, the
one we have denoted as {\bf Case~3}, which could be of some physical
relevance.

\section{$Br(B\rightarrow X_{\lowercase{s}}\gamma)$, $\Delta\rho$, 
$R_{\lowercase{b}}$ and the constrained physical model}
\label{bsg}

As it is the case for 2HDM's without FCNC's (model~I and model~II)
\cite{grant}, so also in model~III it is very difficult to reconcile
the measured value of the inclusive branching fraction for
$B\rightarrow X_s\gamma$ \cite{alam}

\be 
Br(B\rightarrow X_s\gamma)=(2.32\pm 0.51\pm 0.29\pm 0.32)\times
10^{-4}
\label{bsg_exp}
\ee

\noindent with the experimental results for $R_b$, the ratio of the
rate for $Z\rightarrow b\bar b$ to the rate for $Z\rightarrow$hadrons.
The present situation \cite{moriond} is such that $R_b^{\rm
expt}>R_b^{\rm SM}$ ($\sim 3\sigma$) \cite{rb}

\be
R_b^{\rm expt}= 0.2202\pm 0.0016 \,\,\,\,\,,\,\,\,\,\,\,
R_b^{\rm SM}=0.2156 \label{rb_expSM}
\ee

\noindent and the value of $R_b^{\rm expt}$ seems to challenge many
extensions of the SM \cite{bamert,rbrc}. However, several issues on
the measurement of this observable are still unclear and require
further scrutiny \cite{rbrc,isi}. It is plausible that the
experimental situation will change in the future. Therefore we may
want to consider both the case in which the constraint from $R_b^{\rm
expt}$ is enforced and the case in which it is disregarded. A third
crucial electroweak (EW) observable in this analysis is given by the
$\rho$ parameter, which turns out to be very sensitive to the choice
of the mass parameters of any new physics beyond the SM\null. In a
recent paper \cite{rbrc} we have studied the problem in detail,
considering two major scenarios in which the constraint from $R_b$ is
either enforced or disregarded. In the following we will summarize the
main results of Ref.~\cite{rbrc} and discuss them in the context of
the more general picture of model~III that we have been tracing till
here.

Let us first recall how the presence of an extended scalar sector and
of the new FC couplings affects the theoretical prediction for
$B(B\rightarrow X_s\gamma)$, $\Delta\rho$ and $R_b$.

The SM result for the inclusive branching ratio $Br(B\rightarrow
X_s\gamma)$ \cite{ciu_bsg1,buras_bsg} can be modified in order to
include the contributions from the new scalar fields and we obtain
that in model~III

\be
R=\frac{Br(B\rightarrow X_s\gamma)}{Br(B\rightarrow X_ce\bar v_e)}
\sim\frac{\Gamma(b\rightarrow s\gamma)}{\Gamma(b\rightarrow
ce\bar\nu_e)} = \frac{6\alpha}{\pi\,f(m_c/m_b)}\, F\,
\left(|C_7^{(R)}(m_b)|^2+|C_7^{(L)}(m_b)|^2\right) 
\label{bsg_ratio}   
\ee

\noindent where $f(m_c/m_b)$ is the phase-space factor for the
semileptonic decay and $F$ takes into account some $O(\alpha_s)$
corrections to both $B\rightarrow X_c e\bar\nu_e$ and $B\rightarrow
X_s\gamma$ decays (see refs.~\cite{rbrc,ciu_bsg2} for further
comments).  $C_7^{(R)}(m_b)$ and $C_7^{(L)}(m_b)$ are the Wilson
coefficients of the two magnetic type operators that occur in
model~III, for arbitrary $\xi^{U,D}_{ij}$ couplings (see
Ref.~\cite{rbrc}),

\be
Q_7^{(R,L)}=\frac{e}{8\pi^2}m_b\bar s\sigma^{\mu\nu}(1\pm\gamma_5)b 
F_{\mu\nu}
\ee

\noindent in the $\Delta S\!=\!1$ effective hamiltonian evaluated at
the scale $\mu=m_b$. The approximate calculation of $C_7^R(m_b)$ and
$C_7^L(m_b)$ shows that in model~III the $Br(B\rightarrow X_s\gamma)$
is always larger than the SM one \cite{rbrc}. This feature is very
general and the enhancement of the $Br(B\rightarrow X_s\gamma)$
depends on the assumptions we make on the new FC couplings and on the
masses of the neutral and charged scalar fields.

{}From the analysis of Ref.~\cite{rbrc} it is clear that the
$Br(B\rightarrow X_s\gamma)$ is very sensitive to any enhancement of
the FC couplings.  The neutral scalar and pseudoscalar contributions,
involving truly new kind of diagrams with respect not only to the SM
case but also to model~I and model~II 2HDM's, are proportional to
$\xi^D_{bb}$ and $\xi^D_{sb}$.  The charged scalar contribution
depends on the charged couplings $\xi^U_{u_id_j}$ and $\xi^D_{u_id_j}$
for $u_i=c,t$ and $d_j=s,b$ \cite{bsg}.  In particular, the really
leading contribution arises from the diagram with a top quark in the
loop and the relevant couplings will then be: $\xi^{U,D}_{ts}$ and
$\xi^{U,D}_{tb}$. According to Eq.~(\ref{charged}) they are explicitly
given by

\bea
\xi^U_{ts} &=& \xi^U_{tt}V_{ts}+\xi^U_{tc}V_{cs} \,\,\,\,\,\,,
\,\,\,\,\,\,\xi^U_{tb} = \xi^U_{tt}V_{tb}+\xi^U_{tc}V_{cb}\\
\xi^D_{ts} &=& \xi^D_{ss}V_{ts}+\xi^D_{sb}V_{tb} \,\,\,\,\,\,,
\,\,\,\,\,\,\xi^D_{tb} = \xi^D_{sb}V_{ts}+\xi^D_{bb}V_{tb}\,\,\,.
\nonumber
\eea

\noindent Therefore the final prediction for the $Br(B\rightarrow
X_s\gamma)$ will depend on the whole set of FC couplings which involve
the second and the third generation. A strong enhancement of any of
the $\lambda_{ij}$ would conflict with the experimental prediction in
Eq.~(\ref{bsg_exp}), unless some very specific assumptions on the
other parameters (couplings and masses) are made \cite{bsd}.

Let us now consider $R_b$, defined as

\be
R_b \equiv \frac{\Gamma(Z\rightarrow b\bar b)}{\Gamma(Z\rightarrow
\hbox{hadrons})}\,\,\,.
\label{rb_def} 
\ee

\noindent Neglecting all finite quark mass effects ($m_q\sim 0$)
\cite{massless}, the generic expression for $\Gamma(Z\rightarrow q\bar
q)$, for $q=b,\,c,\ldots$, can be written as

\be \Gamma(Z\rightarrow q\bar q)=\frac{N_c}{6}\frac{\alpha_e}
{s_{\sss W}^2  c_{\sss W}^2}M_Z
\left[(\Delta_{q,L})^2+(\Delta_{q,R})^2\right] \label{gammaqq}
\ee

\noindent where $\alpha_e$ is the QED fine structure constant, $N_c$
the number of colors, $s_{\sss W}$ the Weinberg angle and
$\Delta_{q,L(R)}$ the chiral left and right couplings of the $Zq\bar
q$ vertex. They can be written as the sum of a SM piece plus a
correction induced, in our case, by the new FC scalar couplings of
model~III

\be
\Delta_{q,L(R)}\equiv\Delta_{q,L(R)}^{\sss{\rm SM}}+
\Delta_{q,L(R)}^{\sss{\rm NEW}}\,\,\,.
\label{deltadef}
\ee

\noindent Since $R_b$ is in the form of a ratio between two hadronic
widths, most EW oblique and QCD corrections cancel, in the massless
limit, between the numerator and the denominator. The remaining ones,
are absorbed in the definition of the renormalized couplings
$\hat\alpha$ and ${\hat s}_{\sss W}$ (${\hat c}_{\sss W}$), up to
terms of higher order in the electroweak corrections
\cite{pich,kuhn,grant}. As a consequence, the $\Delta_{q,L(R)}$
couplings will be as in Eq.~(\ref{deltadef}), with
$\Delta_{q,L(R)}^{\sss{\rm SM}}$ given by the tree level SM couplings
expressed in terms of the renormalized couplings $\hat\alpha$ and
${\hat s}_{\sss W}$ (${\hat c}_{\sss W}$). This feature makes the
study of $R_b$ particularly interesting, because the new FC
contributions may be easily disentangled in the $Zq\bar q$ vertex
corrections. Using Eqs.~(\ref{gammaqq}) and (\ref{deltadef}), we can
express $R_b$ in terms of $R_b^{\sss{\rm SM}}$ and $R_c^{\sss{\rm
SM}}$ as follows

\be
R_b  =  R_b^{\sss{\rm SM}} \frac{1+\delta_b}{[1+R^{\sss{\rm SM}}_b
\delta_b+R^{\sss{\rm SM}}_c\delta_c]} \label{rqrsmb}
\ee

\noindent where

\be
\delta_q = 2\,\frac{\Delta^{\sss{\rm SM}}_{qL} \Delta^{\sss{\rm
NEW}}_{qL} + \Delta^{\sss{\rm SM}}_{qR} \Delta^{\sss{\rm
NEW}}_{qR}}{(\Delta^{\sss{\rm SM}}_{qL})^2 + ( \Delta^{\sss{\rm
SM}}_{qR})^2}  \label{deltaq}
\ee

\noindent for $q=b,c$ \cite{rurdrs}. In Eqs.~(\ref{rqrsmb}) and
(\ref{deltaq}), terms of $O((\Delta^{\sss{\rm NEW}}_{bL(R)})^2)$ have
been neglected and the numerical analysis confirms the validity of
this approximation.  The vertex corrections $\Delta^{\sss{\rm
NEW}}_{qL}$ and $\Delta^{\sss{\rm NEW}}_{qR}$ in model~III will depend
on the new FC couplings and on the scalar masses. As explained in
Ref.~\cite{rbrc}, the dominant contributions to $\Delta^{\sss{\rm
NEW}}_{qL,R}$ are due to the charged scalar and tend to further
decrease the SM result, pushing the theoretical prediction for $R_b$
in the wrong direction with respect to the current experimental result
in Eq.~(\ref{rb_expSM}). The neutral scalar and pseudoscalar
contributions would increase $R_b$, but they are very small.  In order
to enhance them we have to make very strong requirements on the
couplings and masses of model~III, which correspond to the scenario we
identified as case~3. This scenario and its compatibility with the
other constraints will be discussed later on in this section.

Finally let us consider the $\rho$-parameter, i.e.\ the radiative
corrections to the relation between $M_W$ and $M_Z$. We may expect
that the $W$ and $Z$ propagators are modified by the presence of new
scalar-gauge field couplings (see Appendix~\ref{fr_mod3}). In fact,
the relation between $M_W$ and $M_Z$ is modified by the presence of
new physics and the deviation from the SM prediction is usually
described by introducing the parameter $\rho_0$ \cite{pdg,lang},
defined as

\be
\rho_0=\frac{M_W^2}{\rho M_Z^2\cos^2\theta_W} \label{rhozero_def}
\ee

\noindent where the $\rho$ parameter absorbs all the SM corrections
to the gauge boson self-energies. We recall that the most important SM
correction at the one loop level is induced by the top quark
\cite{grant,lang}

\be
\rho_{\rm top}\simeq\frac{3 G_F m_t^2}{8\sqrt{2}\pi^2}\,\,\,.
\label{rhotop}
\ee

\noindent Within the SM with only one scalar SU(2) doublet
$\rho_0^{\rm tree}=1$. In the presence of new physics we have

\be
\rho_0=1+\Delta\rho_0^{\sss{\rm NEW}} \label{rhozero}
\ee

\noindent where $\Delta\rho_0^{\sss{\rm NEW}}$ can be written in terms
of the new contributions to the $W$ and $Z$ self-energies as

\be
\Delta\rho_0^{\sss{\rm NEW}}=\frac{A^{\sss{\rm NEW}}_{WW}(0)}{M_W^2}-
\frac{A^{\sss{\rm NEW}}_{ZZ}(0)}{M_Z^2}\,\,\,.
\ee

\noindent The determination of $m_t$ from FNAL \cite{cdf} allows us to
distinguish between $\rho_0$ and $\rho\simeq 1+\rho_{\rm top}$. From
the recent global fits of the electroweak data, which include the
input for $m_t$ from Ref.~\cite{cdf} and the new experimental results
on $R_b$, $\rho_0$ turns out to be very close to unity. For
$R_b$=$R^{\rm expt}_b$ as in Eq.~(\ref{rb_expSM}) and $m_t=(174\pm
16)$ GeV, Ref.~\cite{lang} quotes

\be
\rho_0=1.0004\pm 0.0018\pm 0.0018 \,\,\,.
\label{rhozero_gf}
\ee

\noindent This result clearly imposes stringent limits on the
parameters of any extended model. Using the general analytical
expressions in Ref.~\cite{bert}, and adapting the discussion to
model~III (making use of the Feynman rules given in
Appendix~\ref{fr_mod3}), we find that

\be
\Delta\rho_0^{\sss{\rm NEW}}\simeq\frac{G_F}{8\sqrt{2}\pi^2}
\left[\sin^2\!\alpha\,
G(m_c,m_A,m_H)+\cos^2\!\alpha\, G(m_c,m_A,m_h)\right] 
\label{deltarhozero}
\ee

\noindent where all the terms of order $(M_{W,Z}^2/m_c^2)$ have been
neglected and we define

\be
G(m_c,m_A,m_{H,h})= m_c^2-
\frac{m_c^2 m_A^2}{m_c^2-m_A^2}\log{\frac{m_c^2}{m_A^2}}
-\frac{m_c^2 m_{H,h}^2}{m_c^2-m_{H,h}^2}\log{\frac{m_c^2}{m_{H,h}^2}}
+\frac{m_A^2 m_{H,h}^2}{m_A^2-m_{H,h}^2}
\log{\frac{m_A^2}{m_{H,h}^2}} \,\,\,.
\label{gfunction}
\ee

\noindent Therefore the choice of the set of scalar masses will be
crucial in order to make $\Delta\rho_0$ compatible with
Eq.~(\ref{rhozero_gf}).

\medskip
The results of our analysis of model~III \cite{rbrc} indicate that
there are two main available scenarios depending on our choice of
enforcing $R_b$ as additional constraint or not. As far as the
assumptions on the FC couplings are concerned, they correspond to what
in Sec.~\ref{mixing} we called {\bf Case~2} and {\bf Case~3}.
However, new restrictions on the mass parameters have been imposed by
the additional constraints we have discussed in this
section. Therefore we will enlarge the definition of case~2 and case~3
to include also the bounds imposed on the mass parameters. The
resulting two scenarios will constitute the two available physical
{\it solutions} of model~III\null.  We will devote the rest of this
section to illustrate them in detail and summarize their relevant
features.

\begin{enumerate}

\item [{\bf 1.}] If we {\bf enforce the constraint from $R_b^{\rm
expt}$} (see Eq.~(\ref{rb_expSM})), then we can accommodate the
present measurement of the $Br(B\rightarrow X_s\gamma)$ (see
Eq.~(\ref{bsg_exp})) and of the $\Delta F\!=\!2$ mixings (see
Table~\ref{f0f0}) and at the same time satisfy the global fit result
for the $\rho$ parameter (see Eq.~(\ref{rhozero_gf})) provided the
following conditions are satisfied.
\begin{enumerate}
\item[i)] The neutral scalar $h^0$ and the pseudoscalar $A^0$ are very
light, i.e.

\be
50\, \mbox{GeV} \le m_h\sim m_A < 70 \,\mbox{GeV}\,\,.
\label{neutmass}
\ee

\item[ii)] The charged scalar $H^+$ is heavier than $h^0$ and $A^0$,
but not too heavy to be in conflict with the constraints from the
$\rho$ parameter in Eq.~(\ref{rhozero_gf}). Thus

\be
150\, \mbox{GeV}\le m_c\le 200 \,\mbox{GeV} \,\,.
\label{chargmass}
\ee

\item[iii)] The $\xi^D_{ij}$ couplings are enhanced with respect to
the $\xi^U_{ij}$ ones, as described by the pattern we identified as
{\bf Case~3}

\bea 
\label{rbscenario}
\lambda_{bb}&\gg& 1\,\,\,\, \mbox{and}\,\,\,\, \lambda_{tt}\ll 1\\ 
\lambda_{sb}&\gg& 1\,\,\,\, \mbox{and}\,\,\,\, \lambda_{ct}\ll 1\,\,.
\nonumber 
\eea 
\end{enumerate}

\noindent The choice of the phase $\alpha$ is not as crucial as the
above conditions and therefore we do not make any assumption on it.
Within this scenario $R_b$ can be predicted to be less than $2\sigma$
away of $R_b^{\rm expt}$. We refer the reader to Ref.~\cite{rbrc} for
more details. From the previous requirements on the parameter of the
model, we understand that it is in general very difficult to
accommodate the present value of $R_b^{\rm expt}$ in model~III\null.
However, if we assume that the FC couplings (namely, the
$\lambda_{ij}$ parameters) are arbitrary and dictated only by
phenomenology, then it is still possible to find a very small region
of the parameter space in which, in principle, model~III is compatible
with the important experimental results. The values of the neutral
scalar and pseudoscalar masses are required to fit the narrow window
of Eq.~(\ref{neutmass}) and are very close to their experimental lower
bounds. In order to increase them and still agree with $R_b^{\rm
expt}$, we would need a heavier $m_c$ and this would be in conflict
with Eq.~(\ref{chargmass}).

We recall that similar difficulties are present in model~II as well
\cite{grant}. However the important difference with respect to model~II is
that this analysis of the constraints of model~III gives us some 
hints on the possible range of the new FC couplings. This can be used
to explore interesting experimental consequences in FC transitions.
Let us review some of the most important ones.

As we can read in Table~\ref{f0f0}, also in {\bf Case~3} the main
contribution to the mixing comes from the one loop diagrams. Contrary
to the $B^0_d\!-\!\bar B^0_d$ and $K^0\!-\!\bar K^0$ case, we note
that in the $D^0\!-\!\bar D^0$ case the mixing can be much bigger than
in the SM, although still a couple of orders of magnitude below the
experimental bound. This is the only case in which, in a
$F^0\!-\!\bar F^0$ mixing, a loop contribution of model~III can be
much bigger than the SM prediction. Therefore an improved experimental
determination of the $D^0\!-\!\bar D^0$ mixing would be very effective
in constraining the model parameters for case~3.

Moreover the possibility of having large $\xi^D_{bb}$ and $\xi^D_{sb}$
couplings, as predicted by this scenario and allowed by the present
constraint on $B^0_s\!-\!\bar B^0_s$, seems to be particularly
interesting for the study of some rare B-decays (i.e.\ $B\rightarrow
l^+l^-$ and $B\rightarrow X_s l^+l^-$) \cite{savage,hou} and of
$Z\rightarrow \bar bs+b\bar s$. Due to its importance we will discuss
this subject more extensively in Sec.~\ref{b0s}.

Finally, surprisingly enough, we have verified that the cross section
for top-charm production and the decay rate for the rare top decays
that we discussed in Sec.~\ref{tcgZ} are not suppressed even if the
$\xi^U_{ij}$ couplings are. The contribution from the neutral scalars
and pseudoscalar are clearly negligible and the final result is
dominated by the charged scalar contribution. In this case, the
analysis has to be extended with respect to the description we give in
Appendix~\ref{formfactors}, in order to include the complete
expression for the charged couplings (see Eq.~(\ref{charged})). In
fact, the contribution from the charged scalar will be dominated by
the $\hat\xi^D_{\rm charged}$ coupling of Eq.~(\ref{charged}) instead
of by the $\hat\xi^U_{\rm charged}$ ones as we assumed in
Sec.~\ref{eetc}-\ref{tcgZ} and in Appendix~\ref{formfactors}. The
results of Appendix~\ref{formfactors} are still valid, provided some
changes in the $\xi^V$, $\xi^A$, \dots couplings are made.

\item [{\bf 2.}]If we {\bf disregard the constraint from $R_b^{\rm
expt}$} there is no need anymore to impose the bounds of
Eqs. (\ref{neutmass})-(\ref{rbscenario}) and we can safely work in the
scenario of {\bf Case~2}, where only the first generation FC couplings
are suppressed

\be
\lambda_{ui},\lambda_{dj}\ll 1\quad\mbox{for}\quad i,j=1,2,3
\ee

\noindent in order to satisfy the experimental constraints on the
$F^0\!-\!\bar F^0$ mixings. We will assume the FC couplings of the
second an third generations to be given by Eq.~(\ref{coupl_sher}) with
$\lambda_{ij}\!\simeq\!O(1)$. 

\begin{figure}[htb]
\centering
\epsfxsize=4.in
\leavevmode\epsffile{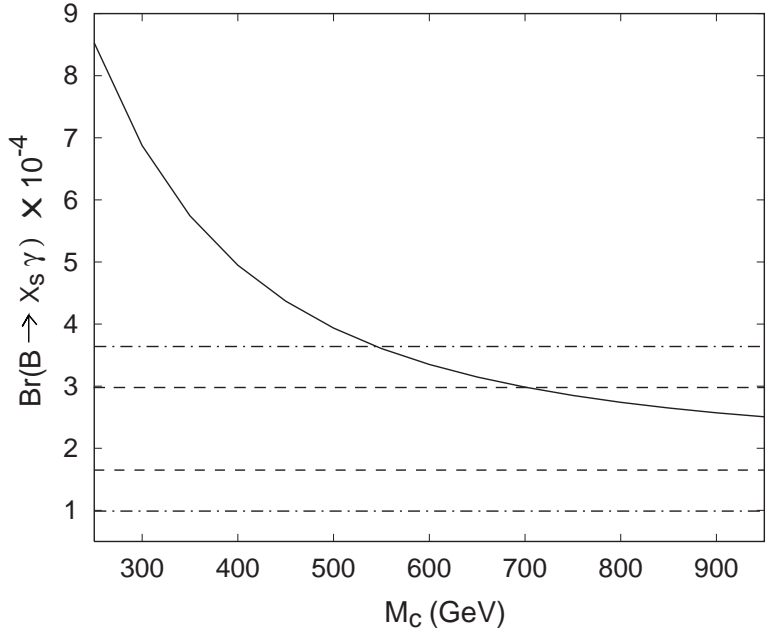}
\caption[]{$Br(B\rightarrow X_s\gamma)$ in model~III\null. The
experimental result at $1\sigma$ (dashed) and $2\sigma$ (dot-dashed)
is also given.}
\label{fig_bsg}
\end{figure}

In this case model~III predicts a $Br(B\rightarrow X_s\gamma)$
compatible with experiments at the $2\sigma$-level \cite{2sigma}, for
$m_c\ge 600$ GeV, as we can see in Fig.~\ref{fig_bsg}. As soon as
$\xi^D_{bb}$ is not enhanced anymore, the contribution of the neutral
scalars and pseudoscalar is completely negligible. Therefore, both the
value of the mixing angle $\alpha$ and of the neutral scalar and
pseudoscalar masses ($m_{H}$, $m_h$, and $m_A$) are irrelevant. In
particular, Fig.~\ref{fig_bsg} is obtained for $\alpha=0$ and values
for $m_h$ and $m_A$ resulting from the fit to $\Delta\rho_0$
\cite{rbrc}

\be
m_H,m_h\le m_c\le m_A\,\,\,\,\,\mbox{and}\,\,\,\,
m_A\le m_c\le m_H,m_h \,\,\,.
\label{Mc_inbetween}
\ee

\noindent We note that none of the previous scenarios would give an
enhanced value of $R_b$, because in that case $m_A$ and $m_h$ would be
required to be equal and light (see Eq.~(\ref{neutmass})). In fact, in
this scenario model~III predicts $R_b$ to be slightly smaller than
$R_b^{\rm SM}$ (and $R_c$ slightly bigger than $R_c^{\rm SM}$)
\cite{rbrc}. 

The main theoretical predictions for case~2 have already been
discussed in many sections of this paper: top-charm production in
Sec.~\ref{eetc}, rare top decays in Sec.~\ref{tcgZ} and
$F^0\!-\!\bar F^0$ mixings in Sec.~\ref{mixing}. 

\end{enumerate}

\section{$B^0_{\lowercase{s}}\!-\!\bar B^0_{\lowercase{s}}$, 
$Z\rightarrow \lowercase{\bar b}\lowercase{s}$ and some rare B-decays}
\label{b0s}

We have seen in Sec.~\ref{bsg} that the present experimental
measurement of $R_b$ (see Eq.~(\ref{rb_expSM})) suggests a new pattern
of FC couplings for model~III, that we called {\bf Case~3}, according
to which the $\xi^D_{ij}$ couplings would be enhanced with respect to
the $\xi^U_{ij}$ ones.  In this section we want to examine the
compatibility of this assumption with the $B^0_s\!-\!\bar B^0_s$
mixing and its phenomenological implications for the rare decays:
$B^0\rightarrow X_s \mu^+\mu^-$, $B_s\rightarrow \mu^+\mu^-$ and
$Z\rightarrow bs$ (we denote in this way the sum of the final states
$\bar b s$ and $b\bar s$).

We have summarized our present theoretical and experimental knowledge
of these processes in Table~\ref{bs}, where the predictions of case~3
of model~III are compared with the SM results and with the results of
the 2HDM's with natural flavor conservation (model~I and model~II). The
experimental situation is still uncertain for all of the processes
under study and we report in Table~\ref{bs} the existing experimental
bounds.  It is
clear from Table~\ref{bs} that in these processes there are good
chances that continued experimental search could show deviations from the
SM predictions.

As is the case for the other $\Delta F\!=\!2$ mixings that we have
examined in Sec.~\ref{mixing}, $B^0_s\!-\!\bar B^0_s$ is extremely
important to constrain the corresponding FC coupling,
i.e.\ $\xi^D_{sb}$. However, for the time being the experiments give us
only a lower bound \cite{aleph}

\be \Delta M_{B_s} > 4.3\times 10^{-12}
\,\mbox{GeV}\quad\mbox{or}\quad x_s\equiv\frac{\Delta M_{B_s}}
{\Gamma_{B_s}}>10 \,\,\,.
\label{B0s_bound}
\ee

\noindent Specifying Eq.~(\ref{f0f0_tree}) to the $B^0_s$ case, from
Eq.~(\ref{B0s_bound}) we get a lower bound for the $\xi^D_{sb}$ FC
coupling.  Therefore $B^0_s\!-\!\bar B^0_s$ tells us that the
$\xi^D_{sb}$ coupling need not be small, i.e.\ $\lambda_{sb}$ can be
somewhat bigger than one. We know from the analysis of Sec.~\ref{bsg}
that this compatibility is realized in case~3 of
model~III\null. Moreover, we have verified that the presence of an
enhanced $\xi^D_{sb}$ coupling does not represent a problem for
non-leptonic decays of the $b$ quark, in particular for those decays
which arise at the tree level via $b\rightarrow sc\bar c$.  In the
spirit of this qualitative analysis, we ask that at the quark level
the contribution from new physics to the rate $\Gamma(b\rightarrow
sc\bar c)$ is appreciably smaller than the corresponding SM one. This
is indeed the case for $\lambda_{sb}<40$ , i.e.\ in a large range of
values.

It is remarkable that the $B^0_s\!-\!\bar B^0_s$ mixing does not
prevent $\xi^D_{sb}$ from belonging to that small region of the
parameter space of model~III which is suggested by $R_b^{\rm
expt}$. Therefore we want to further investigate the phenomenological
consequences of case~3 of model~III in some physical processes that
more directly involve the $\xi^D_{sb}$ coupling: the rare $B$-decays
$B\rightarrow X_s \mu^+ \mu^-$ and $B^0_s\rightarrow \mu^+ \mu^-$ and
the decay $Z\rightarrow bs$.

\begin{table}
\begin{center}
\begin{tabular}{c c c c c}
\\ Process & SM & \parbox{2cm}{\centering 2HDM's Model I,II} &
\parbox{2cm}{\centering Model III Case 3} & Experiment\\ \\ \hline\\
$\Delta M_{B_s}$ & $10^{-12}\!\!-\!\!10^{-11}$ & $10^{-12}\!\!-\!\!10^{-10}$ &
$>9\cdot 10^{-12}$ & $>4.3\cdot 10^{-12}$ \cite{aleph} \\ \\
$Br(B^0_s\rightarrow \mu^+\mu^-$)& $\sim 4\cdot 10^{-9}$ & 
$10^{-9}\!-\!10^{-8}$ &
\parbox{2.5cm}{\centering $10^{-7}\!-\!10^{-4}$
($\lambda_{\mu\mu}\simeq 1$)} & \parbox{2.8cm}{\centering 
$<8.3\cdot 10^{-6}$ \cite{ua1_bs} $<8.4\cdot 10^{-6}$ \cite{cdf_bs}}\\ \\
$Br(B^0\rightarrow X_s\mu^+\mu^-)$ & $\sim 7\cdot 10^{-6}$ &
$10^{-5}\!-\!10^{-4}$ & \parbox{2.5cm}{\centering $10^{-6}\!-\!10^{-4}$ 
($\lambda_{\mu\mu}\simeq 1$)} & \parbox{2.8cm}{\centering 
$<2.5\cdot 10^{-5}$ \cite{cdf_bs} $<5\cdot 10^{-5}$ \cite{ua1}} \\ \\
$Br(Z\rightarrow bs)$ & $\sim 6\cdot 10^{-8}$ & $\sim 10^{-8}$ & 
$10^{-8}-10^{-6}$ & ? \\ \\
\end{tabular}
\caption[]{Values of $\Delta M_{B_s}$ (GeV),
$Br(B_s\rightarrow\mu^+\mu^-)$, $Br(B_s\rightarrow X_s\mu^+\mu^-)$,
and $Br(Z\rightarrow bs)$ in the SM, in 2HDM's with natural flavor
conservation (Model~I and Model~II) and in case~3 of
Model~III\null. Each range is obtained by varying the parameters of
the corresponding model over a large region of the parameter space,
compatible with phenomenology and with the assumption of weakly
coupled scalar fields.The present experimental bounds are also
given. The top mass is taken to be $m_t\simeq 180$ GeV.}
\label{bs}
\end{center}
\end{table}

The phenomenological relevance of the decay $B^0_{d,s}\rightarrow
l^+l^-$ and in particular of $B_{d,s}\rightarrow \mu^+\mu^-$ has been
pointed out in Ref.~\cite{savage}. Although the experimental
measurement is still poor \cite{ua1_bs,cdf_bs}, this is a rare but
theoretically very clean B-decay, which is not affected by large QCD
corrections. In model~III it can arise at the tree level via the
exchange of a neutral scalar or pseudoscalar with FC interactions.  We
think that the possibility of having a large $\xi^D_{sb}$ coupling
prompts us to reconsider $B^0_s\rightarrow \mu^+\mu^-$. As is the case
for $B^0_d$, the prediction for $B^0_s\rightarrow \mu^+\mu^-$ in
model~III can be enhanced at least by a factor of $10^2$ with respect
to the SM and to the 2HDM's with natural flavor conservation. However,
the range reported in Table~\ref{bs} is obtained for a $\xi_{\mu\mu}$
coupling given by Eq.~(\ref{coupl_sher}) with
$\lambda_{\mu\mu}\!\simeq\! 1$ and different enhancements of the
$\xi^D_{sb}$ coupling. As we said from the very beginning, we do not
want to consider here the implementation of model~III in the leptonic
sector. Therefore we will take the number of Table~\ref{bs} just as an
indication of the possibility of new interesting signals from this
rare decay.

The $B\rightarrow X_s \mu^+\mu^-$ case could be even more interesting.
In fact, as we can see from Table~\ref{bs}, better experimental bounds
\cite{cdf_bs,ua1} exist and are nowadays only one order of magnitude
away from the SM prediction, which is known to very high accuracy
(including QCD corrections, long distance effects, etc.)
\cite{grinstein,buras_bsll}. Therefore this decay could become a good
constraint for the $\xi^D_{bb}$ and $\xi^D_{sb}$ couplings. In
model~III there are two possible contributions at the parton level: a
one loop transition due to the one loop induced effective $Zbs$ vertex
and a tree level transition directly mediated by a non-standard
neutral scalar or pseudoscalar ($H^0$, $h^0$ or $A^0$). The tree level
contribution crucially depends on the order of magnitude of the
$\xi^D_{sb}$ coupling and on the assumption we make on the
$\xi_{\mu\mu}$ coupling. In particular, if $\xi^D_{sb}$ is enhanced
and $m_h,m_A$ are not too heavy (as in case~3), this contribution
could become very important for $\xi_{\mu\mu}$ given by
Eq.~(\ref{coupl_sher}) with $\lambda_{\mu\mu}\!\simeq\!O(1)$.  However
the dependence on $\xi_{\mu\mu}$ does not allow us to make strong
statements. As we explained before, the one loop contribution in
model~III does not depend on $\xi_{\mu\mu}$ and can be comparable, for
large $\xi^D_{bb}$ and $\xi^D_{sb}$, to the SM contribution.  Indeed
the effective $Zbs$ vertices in the SM and in model~III are given
respectively by

\bea
(V_{Zbs}^\mu)_{\rm SM}&\simeq & \frac{g}{c_W}\frac{\alpha_e}{4\pi
s_W^2}(\Delta_L^{bs})_{\rm SM} \bar b_L\gamma^\mu s_L \\
\label{Zbs_vertex}
(V_{Zbs}^\mu)_{\rm MOD_3}&\simeq & \frac{g}{c_W}\frac{\alpha_e}{4\pi
s_W^2} [(\Delta_L^{bs})_{\rm MOD_3} \bar b_L\gamma^\mu s_L+
(\Delta_R^{bs})_{\rm MOD_3} \bar b_R\gamma^\mu s_R]\nonumber
\eea

\noindent where $\alpha_e$ is the QED fine structure constant, $s_W$
the Weinberg angle and we have used the notation
$\gamma^\mu_{L,R}=(1\pm\gamma^5)/2$.  Comparing the
$\Delta_{L,R}^{bs}$ couplings we note that $(\Delta_L^{bs})_{\rm
SM}\!\simeq\!0.1$ while $(\Delta_{L,R}^{bs})_{\rm MOD_3}$ depend on
$\xi^D_{bb}$ and $\xi^D_{sb}$. For instance, when $\lambda_{bb}\simeq
40$ and $\lambda_{sb}\simeq 10$, $(\Delta_L^{bs})_{\rm MOD_3}\simeq
0.3$, $(\Delta_R^{bs})_{\rm MOD_3}\simeq 0.06$ and the contribution of
model~III to the $Br(B\rightarrow X_s\mu^+\mu^-)$ becomes comparable
or even larger than the SM one.  Therefore, this decay can play an
important role in confirming the case~3 scenario of model~III and the
compatibility of the model with the present experimental prediction
for $R_b$ and the most important other EW constraints.

\begin{figure}
\centering
\epsfxsize=4.in
\leavevmode\epsffile{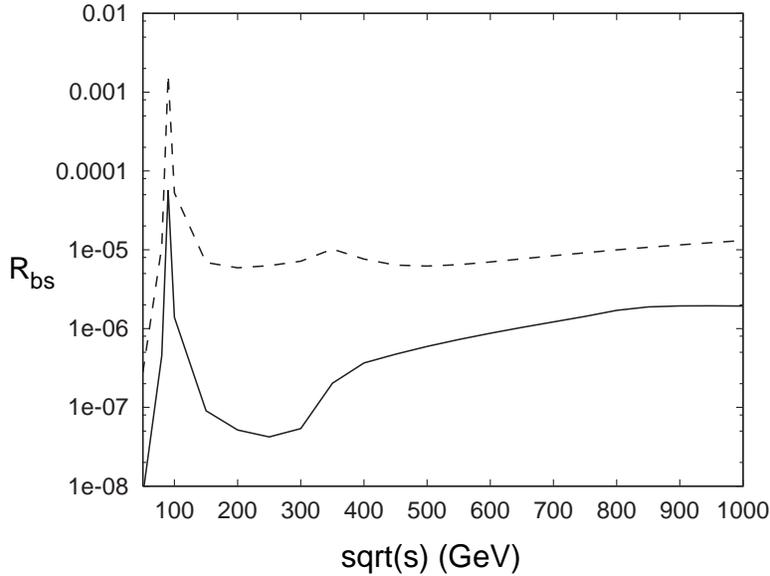}
\caption[]{$R^{bs}$ as a function of $\sqrt s$ for the two scenarios
of case 2 (solid) and case 3 (dashed).}
\label{epembs}
\end{figure}

Finally, the effective $Zbs$ vertex also induces a non negligible rate
for $Z\rightarrow bs$. This rate is given in the SM and in model~III
respectively by

\bea
\Gamma(Z\rightarrow bs)_{\rm SM}&\simeq & 
N_c\frac{\alpha_e}{6s_W^2c_W^2}M_Z
\left(\frac{\alpha_e}{4\pi s_W^2}\right)^2(\Delta_L^{bs})_{\rm SM}^2 \\
\label{Zbs_rate}
\Gamma(Z\rightarrow bs)_{\rm MOD_3}&\simeq & 
N_c\frac{\alpha_e}{6s_W^2c_W^2}M_Z
\left(\frac{\alpha_e}{4\pi s_W^2}\right)^2
\left[(\Delta_L^{bs})^2_{\rm MOD_3}+
(\Delta_R^{bs})^2_{\rm MOD_3}\right]\,\,\,. \nonumber
\eea

In the scenario of case~3, $\Gamma(Z\rightarrow bs)_{\rm MOD_3}$ can
be almost an order of magnitude larger than $\Gamma(Z\rightarrow
bs)_{\rm SM}$. This is reported in terms of branching ratio in Table
\ref{bs}, where the previous rates have been normalized to the $Z$
width ($\Gamma_Z=2.49\pm 0.007$ GeV). Much of the relevance of this
decay mode in model~III depends on the enhancement of the $\xi^D_{ij}$
couplings. Therefore, any experimental bound would be extremely
effective.  We illustrate in Fig.~\ref{epembs} the cross section for
the related process $e^+e^-\rightarrow \bar sb+s\bar b$ normalized to
the cross section for $e^+e^-\rightarrow\gamma^*\rightarrow
\mu^+\mu^-$, i.e.

\be
R^{bs} \equiv \frac{\sigma(e^+e^-\rightarrow b\bar s+ \bar bs)}
{\sigma( e^+e^-\rightarrow \gamma^*\rightarrow \mu^+\mu^-)}
\label{rbs_ee}
\ee

\noindent starting from values of the center of mass energy below the
$Z$-peak up to $\sqrt s\!=\!1$ TeV\null. The upper curve corresponds
to case~3 and the lower curve to case~2.  As we can see, there is a
considerable enhancement in the scenario of case~3 that we are
considering in this section.  Although the $Z\rightarrow bs$ events
are not as distinctive as the $tc$-production events,
Fig.~\ref{epembs} seems to suggest that the experimental situation may
be favorable and more experimental effort in this direction appears
very worthwhile.

\section{Outlook and conclusions}
\label{end}

In this paper and in Refs. \cite{eetc,mumutc,rbrc} we examined the
phenomenology of FCSC's that may occur in extended models. We strongly
share the point of view with many that the extraordinary mass scale of
the top quark should prompt us all to reexamine our theoretical
prejudices against the existence of such currents, especially
involving the top quark.

A very simple extension of the SM with another Higgs doublet leads
rather naturally to such scalar currents. The model has the nice
feature that experimental information can be systematically catalogued
and guidance for further effort can be sought. The model has important
bearings for some key reactions: $e^+e^-(\mu^+\mu^-)\rightarrow \bar t
c+t\bar c, c\bar c$; $t\rightarrow c \gamma (g,Z)$; $D^0$-$\bar D^0$,
and $B^0_s$-$\bar B^0_s$ oscillations; $B(B_s)\rightarrow l^+l^-$,
$B(B_s)\rightarrow l^+l^-X_s$, and $e^+e^-(Z)\rightarrow b\bar s+\bar
b s$.  Continued experimental effort towards these can hardly be over
emphasized. The model also has important implications for
$Z\rightarrow b\bar b$ and we want to stress that it is extremely
important to clarify the experimental situation regarding
$Z\rightarrow b\bar b$.

\section*{Acknowledgments}

This research was supported in part by U.S. Department of Energy
contracts DC-AC05-84ER40150 (CEBAF) and DE-AC-76CH0016 (BNL).

\newpage
\appendix
\section{Feynman rules for model~III}
\label{fr_mod3}

In this appendix we summarize the Feynman rules for model~III, which
are used in the calculations presented in the paper. We choose to work
in the 't Hooft-Feynman gauge.

\subsection*{Fermion-Scalar couplings}

We present the Feynman rules for the couplings of the scalar fields
$H^1$ (neutral scalar), $H^2$ (neutral pseudoscalar), and $H^+$
(charged scalar), to up-type and down-type quarks, as can be derived
from the Yukawa Lagrangian of model~III
(Eqs.~(\ref{lyukmod3})--(\ref{lyukfc})). Following the discussion of
Sec.~\ref{model}, these are the Feynman rules we need in our
calculation of $R_b$.

\begin{tabular}{p{4cm} p{8.2cm}}\\ \\ \\
\parbox[b]{4cm}{\epsffile{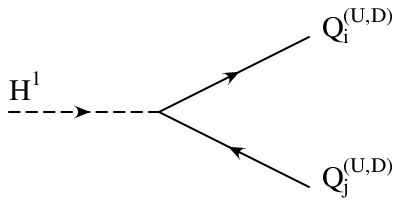}} & 
\raisebox{5.ex}{$\frac{-i}{2\sqrt{2}}\left[(\xi_{ij}^{U,D}+
\xi_{ji}^{U,D*})+(\xi_{ij}^{U,D}-\xi_{ji}^{U,D*})\gamma_5\right]$}
\\ \\ \\
\parbox[b]{4cm}{\epsffile{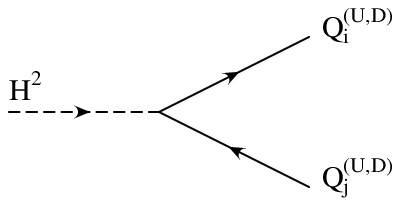}} & 
\raisebox{5.ex}{$\frac{1}{2\sqrt{2}}\left[(\xi_{ij}^{U,D}-
\xi_{ji}^{U,D*})+(\xi_{ij}^{U,D}+\xi_{ji}^{U,D*})\gamma_5\right]$}
\\ \\ \\
\parbox[b]{4.cm}{\epsffile{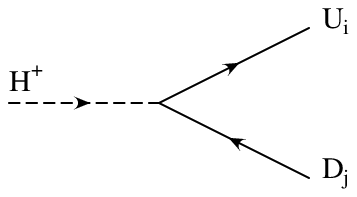}} & 
\raisebox{5.ex}{$\frac{-i}{2}\left[V_{\sss{\rm  CKM}}\!\cdot\!
\xi_{ij}^{D}(1+\gamma_5)-\xi_{ij}^{U}\!\cdot\! V_{\sss{\rm CKM}}
(1-\gamma_5)\right]$}\\ \\ \\
\end{tabular}

Although the $\xi_{ij}^{U,D}$ couplings are left complex in the above,
in practice, in our calculation we assumed they are real,
i.e.\ $\xi_{ij}^{U,D}\simeq\xi_{ij}^{U,D*}$, as we were not concerned
with any phase-dependent effects.

\subsection*{Gauge boson-Scalar couplings}

Here is a list of the $Z$-boson, $W$-boson, and $\gamma$ interactions
with model~III scalar fields. We report them in terms of scalar mass
eigenstates, $\bar H^0$, $h^0$, $A^0$, and $H^+$. We always have to
take note of the relations (see Eqs.~(\ref{masseigen}) and
(\ref{nomasseigen})) between the scalar mass eigenstates and ($H^0$,
$H^1$, $H^2$, $H^+$) and use the fact that neither $Z H^0 H^1$ nor $Z
H^0 H^2$ couplings are present \cite{bert,knowles}. We note the
absence at the tree level of vertices like $A^0 Z^\mu Z^\nu$ and $A^0
W_+^\mu W_-^\nu$.

\begin{tabular}{p{6cm} p{5cm}}\\ \\ \\
\parbox[b]{6cm}{\epsffile{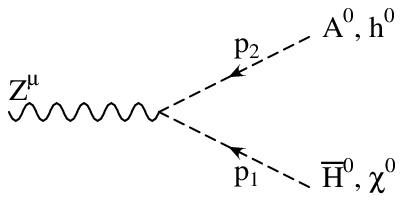}} & 
\raisebox{5.ex}{$\frac{g_{\sss W}}{2c_{\sss W}}\sin\alpha\,
(p_2-p_1)^{\mu}$}\\ \\ \\
\parbox[b]{6cm}{\epsffile{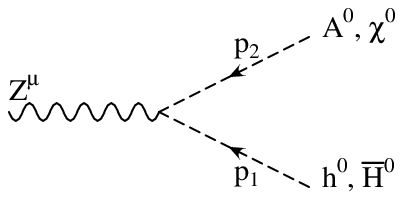}} & 
\raisebox{5.ex}{$\frac{g_{\sss W}}{2c_{\sss W}}\cos\alpha\,
(p_2-p_1)^{\mu}$}\\ \\ \\
\parbox[b]{6cm}{\epsffile{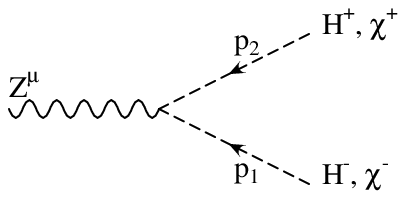}} & 
\raisebox{5.ex}{$\frac{ig_{\sss W}}{2c_{\sss W}}(1-2s_{\sss W}^2)
(p_2-p_1)^{\mu}$}\\ \\ \\
\parbox[b]{6cm}{\epsffile{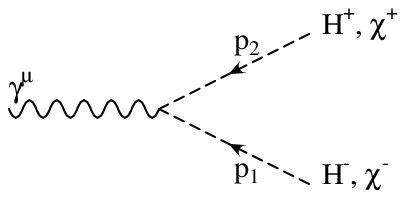}} & 
\raisebox{5.ex}{$ie\,(p_2-p_1)^{\mu}$}\\ \\ \\
\parbox[b]{6cm}{\epsffile{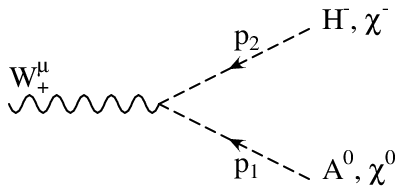}} & 
\raisebox{5.ex}{$\frac{g_{\sss W}}{2}(p_2-p_1)^{\mu}$}\\ \\ \\
\end{tabular}

\begin{tabular}{p{6cm} p{5cm}}\\ \\ \\
\parbox[b]{6cm}{\epsffile{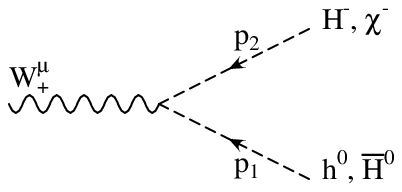}} & 
\raisebox{5.ex}{$\frac{-ig_{\sss W}}{2}\cos\alpha\,(p_2-p_1)^{\mu}$}\\ \\ \\
\parbox[b]{6cm}{\epsffile{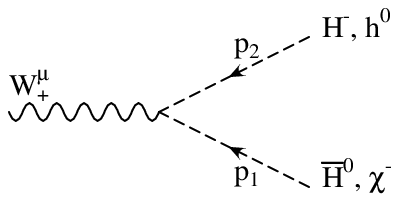}} & 
\raisebox{5.ex}{$\frac{-ig_{\sss W}}{2}\sin\alpha\,(p_2-p_1)^{\mu}$}\\ \\ \\
\parbox[b]{6cm}{\epsffile{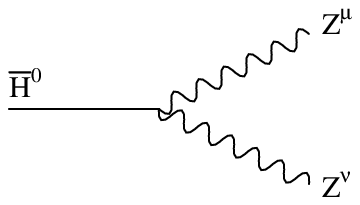}} & 
\raisebox{5.ex}{$i\frac{g_{\sss W}}{c_{\sss W}}M_Z\cos\alpha\,
g^{\mu\nu}$}\\ \\ \\
\parbox[b]{6cm}{\epsffile{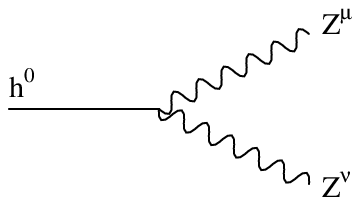}} & 
\raisebox{5.ex}{$-i\frac{g_{\sss W}}{c_{\sss W}}M_Z\sin\alpha\,
g^{\mu\nu}$}\\ \\ \\
\parbox[b]{6cm}{\epsffile{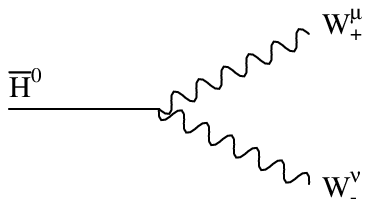}} & 
\raisebox{5.ex}{$ig_{\sss W}M_W\cos\alpha\,g^{\mu\nu}$}\\ \\ \\
\parbox[b]{6cm}{\epsffile{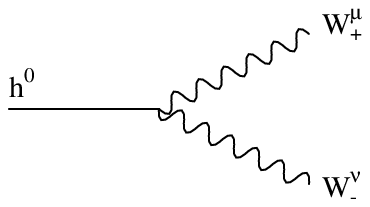}} & 
\raisebox{5.ex}{$-igM_W\sin\alpha\,g^{\mu\nu}$}\\ \\ \\
\end{tabular}

\section{One loop form factors for the $Z\lowercase{tc}$ and the
$\gamma\lowercase{tc}$ vertices}
\label{formfactors}

In this appendix we give the complete analytical expressions of the
one-loop form factors for the $Ztc$, $\gamma tc$, and $gtc$
($g\!=\!$ gluon) vertices, defined by

\be
\Delta^{(V)}_{tc}= \frac{1}{16\pi^2}\bar c\left(A^{(V)}\gamma^{\mu}+
B^{(V)}\gamma^{\mu}\gamma^5+
iC^{(V)}\sigma^{\mu\nu}\frac{q_{\nu}}{m_t}+
iD^{(V)}\sigma^{\mu\nu}\frac{q_{\nu}}{m_t}\gamma^5\right)
tV_{\mu} 
\label{1Lverteces}
\ee

\noindent where $V=\gamma,Z,g$ and in the gluon case we understand
that $g=g^aT^a$ for $a=1,\ldots 8$, $T^a$ denoting the SU(3) color
matrices. This parameterization is obtained by reducing the most
general tensorial form of the $\Delta^{(V)}_{tc}$ vertex using 
the Gordon's decomposition and the gauge invariance of the currents
involved. 

Referring to the definition of the mass scalar eigenstates in
Eq.~(\ref{masseigen}), and using the Feynman rules of
Appendix~\ref{fr_mod3}, we have computed corrections both for
$\alpha\!=\!0$ and for $\alpha\!\neq\! 0$~. At the qualitative level
of our analysis this makes a minor difference. Therefore, we decide to
give in the following the analytical results in the case of
$\alpha\!=\!0$.  In model~III with $\alpha\!=\!0$, $H^0$, $H^1$ and
$H^2$ coincide with the mass eigenstates $\bar H^0$, $h^0$, and $A^0$
and all the new FC contributions come from the second
doublet. Therefore the form factors $A^{(\gamma,Z,g)}$,
$B^{(\gamma,Z,g)}$, \ldots are calculated by summing up all the one loop
corrections generated by the neutral scalar $h^0$, the neutral
pseudoscalar $A^0$ and the charged scalar $H^+$. Following
Ref.~\cite{lukesavage}, we will write each single form factor as the
sum of four different contributions, i.e.

\be
F^{(V)}=F^{(V)}_h+F^{(V)}_A+F^{(V)}_M+F^{(V)}_C \,\,\,\,\,\mbox{for}
\,\,\,\, F=A,B,C,D
\label{FV}
\ee

\noindent where $F^{(V)}_h$ denotes the contribution coming from those
diagrams in which an $h^0$ neutral scalar is exchanged, $F^{(V)}_A$
and $F^{(V)}_C$ the same for the neutral pseudoscalar $A^0$ and for
the charged scalar $H^+$ and finally $F^{(V)}_M$ represents the mixed
$h^0$-$A^0$ contribution (see Fig.~\ref{Zgtc_1loop}).

The choice of $\alpha=0$ and of the previous notation should also help
the comparison with an analogous calculation reported in
Ref.~\cite{lukesavage}, in which the authors computed the
$Br(t\rightarrow c Z)$ and the $Br(t\rightarrow c\gamma)$. Indeed, in
model~III, the effective one loop couplings $Ztc$ and $\gamma tc$
(calculated for $q^2\ne 0$) enter in the same way both in the
calculation of the rates $\Gamma(t\rightarrow cZ,\gamma)$ and in the
cross section for $e^+e^-\rightarrow\gamma^*,Z^*\rightarrow t\bar c
+\bar t c$.  We find that our results are often different from the
ones reported in Ref.~\cite{lukesavage}. In particular, we confirm the
results for $F^{(V)}_h$ and $F^{(V)}_C$, whereas we have different
analytical results for $F^{(V)}_A$ and $F^{(V)}_M$.

As a proof of consistency of our results, we have checked that the sum
of different sets of diagrams is divergence free and that the final
one-loop vertices satisfy the right Ward identities.  We recall that,
for each different form factor $F^{(V)}$ (for $F=A,B,C,D$), the pole
cancellation is verified separately for
$F^{(V)}_h+F^{(V)}_A+F^{(V)}_{M}$, and $F^{(V)}_C$.  In this respect
we note that if we use the results reported in Ref.~\cite{lukesavage},
there is one case in which the cancellation of the divergent terms
among different diagrams seems not to be verified, namely the term
$A^Z_h+A^Z_A+A^Z_M$ in the $A^Z$ form factor \cite{poles}.

We finally note that in the analytical computation of the Feynman
diagrams, all the quarks except the top quark are taken to be
massless. Non-zero light quark masses are kept only in the
$\xi^{U,D}_{ij}$ couplings, according to the assumption made in
Eq.~(\ref{coupl_sher}). 

Our results for the form factors are described in the following,
where, for the sake of comparison, we adopt the notation of
Ref.~\cite{lukesavage}. The form factors are indicated as $F_h,\ldots$
(for $F=A,B,C,D$), where the upper index is dropped in order to
slightly simplify the notation of Eqs.~(\ref{1Lverteces}) and
(\ref{FV}). The distinction between the different vector boson
($\gamma$, $Z$ or $g$) is made at the level of the couplings.

\subsection*{Contribution of $h^0$}

\bea A_h&=&\int_0^1
dx\left[-(1-x)\log\frac{\beta^h}{\mu^2}+\log{\gamma^h}\right]
\left(\xi^h_V\alpha_V-\xi^h_A\alpha_A\right)\nonumber\\ 
&+&\int_0^1\int_0^{1-x}dxdy\left[\left(1+\log\frac{\eta^h}{\mu^2}-
\frac{xyq^2}{\eta^h}\right)\left(\xi^h_V\alpha_V+\xi^h_A\alpha_A
\right)+\frac{m_t^2(x+y-2)}{\eta^h}
\left(\xi^h_V\alpha_V-\xi^h_A\alpha_A\right)\right]\nonumber\\
B_h&=&\int_0^1
dx\left[-(1-x)\log\frac{\beta^h}{\mu^2}+\log{\gamma^h}\right]
\left(\xi^h_V\alpha_A-\xi^h_A\alpha_V\right) \\ 
&+&\int_0^1\int_0^{1-x}
dxdy\left[\left(-1-\log\frac{\eta^h}{\mu^2}+\frac{xyq^2}{\eta^h}
\right)\left(\xi^h_V\alpha_A+\xi^h_A\alpha_V\right)+\frac{m_t^2(x+y-2)}
{\eta^h}\left(\xi^h_V\alpha_A-\xi^h_A\alpha_V\right)\right]\nonumber\\
C_h&=&m_t^2\int_0^1\int_0^{1-x}
 dxdy\left[-\frac{x}{\eta^h}\left(\xi^h_V\alpha_V-
\xi^h_A\alpha_A\right)-\frac{y(2-x-y)}{\eta^h}\left(\xi^h_V\alpha_V+
\xi^h_A\alpha_A\right)\right]\nonumber \\ 
D_h&=&m_t^2\int_0^1\int_0^{1-x}
dxdy\left[-\frac{x}{\eta^h}\left(\xi^h_A\alpha_V-
\xi^h_V\alpha_A\right)-\frac{y(2-x-y)}{\eta^h}\left(\xi^h_V\alpha_A+
\xi^h_A\alpha_V\right)\right]\nonumber \eea

\noindent where we have introduced the following definitions

\bea
\beta^h&=& xm_t^2+(1-x)m_h^2 \nonumber \\
\eta^h&=&(x+xy+y^2)m_t^2+(1-x-y)m_h^2-xyq^2 \\
\gamma^h&=&\frac{xm_t^2+(1-x)m_h^2}{x^2m_t^2+(1-x)m_h^2}\,\,\,,
\nonumber 
\eea

\noindent denoting by $q^2=s$ the mass of the gauge boson (physical
mass or invariant mass depending on the process considered) involved
in the top-charm production process. The couplings $\xi^h_V$ and
$\xi^h_A$ are defined as the following two linear combinations of the
original $\xi^U_{ct}$ and $\xi^U_{tt}$ couplings

\be 
\xi^h_V=\frac{1}{4}\xi^U_{tt}\left(\xi^U_{ct}+\xi^{U*}_{tc}\right)
\,\,\,\,\,\mbox{and}\,\,\,\,\,
\xi^h_A=\frac{1}{4}\xi^U_{tt}\left(\xi^U_{ct}-\xi^{U*}_{tc}\right) 
\ee

\noindent while $\alpha_V=\alpha_V^{\gamma,Z,g}$ and
$\alpha_A=\alpha_A^{\gamma,Z,g}$ denote the vector and axial-vector
couplings of the different gauge bosons, given respectively by

\bea
\label{axialvector_coupl}
\alpha_V^\gamma&=&\frac{2}{3}e\,\,\,\,\,\, , \,\,\,\,\,\, 
\alpha_A^\gamma=0 \\
\alpha_V^Z&=&\frac{g_{\sss W}}{4\cos\theta_W}\left(1-\frac{8}{3}
\sin\theta_{\sss W}\right)\,\,\,\,\,\, , \,\,\,\,\,\, 
\alpha_A^Z=-\frac{g_{\sss W}}{4\cos\theta_{\sss W}}\nonumber\\
\alpha_V^g&=&g_s\,\,\,\,\,\, , \,\,\,\,\,\, 
\alpha_A^g=0 \nonumber
\eea

\noindent where $e$, $g_W$ and $g_s$ are the electromagnetic, weak and
strong coupling constant, while $\theta_W$ is the Weinberg angle.
These results are in agreement with those of
Ref.~\cite{lukesavage}. The apparent difference in sign for $C_h$ and
$D_h$ is only due to a different assumption on the momentum of the
gauge boson. 

\subsection*{Contribution of $A^0$}

\bea
A_A&=&\int_0^1 dx\left[-(1-x)\log\frac{\beta^A}{\mu^2}-\log{\gamma^A}
\right]\left(\xi^A_A\alpha_V-\xi^A_V\alpha_A\right)\nonumber \\
&-&\int_0^1\int_0^{1-x} 
dxdy\left[\left(-1-\log\frac{\eta^A}{\mu^2}+\frac{xyq^2}{\eta^A}
\right)\left(\xi^A_A\alpha_V+\xi^A_V\alpha_A\right)+\frac{m_t^2(x+y)}
{\eta^A}\left(\xi^A_A\alpha_V-\xi^A_V\alpha_A\right)\right]\\
B_A&=&\int_0^1 dx\left[-(1-x)\log\frac{\beta^A}{\mu^2}-\log{\gamma^A}\right]
\left(-\xi^A_V\alpha_V+\xi^A_A\alpha_A\right)\\
&-&\int_0^1\int_0^{1-x} 
dxdy\left[\left(1+\log\frac{\eta^A}{\mu^2}-\frac{xyq^2}{\eta^A}
\right)\left(\xi^A_V\alpha_V+\xi^A_A\alpha_A\right)+\frac{m_t^2(x+y)}
{\eta^A}\left(-\xi^A_V\alpha_V+\xi^A_A\alpha_A\right)\right]\nonumber\\
C_A&=&m_t^2\int_0^1\int_0^{1-x} 
dxdy\left[\frac{x}{\eta^A}\left(\xi^A_A\alpha_V-
\xi^A_V\alpha_A\right)+\frac{y(x+y)}{\eta^A}\left(\xi^A_A\alpha_V+
\xi^A_V\alpha_A\right)\right]\nonumber \\
D_A&=&m_t^2\int_0^1\int_0^{1-x} 
dxdy\left[\frac{x}{\eta^A}\left(\xi^A_V\alpha_V-
\xi^A_A\alpha_A\right)+\frac{y(x+y)}{\eta^A}\left(\xi^A_V\alpha_V+
\xi^A_A\alpha_A\right)\right]\nonumber
\eea

\noindent where, in analogy with the previous case, we have introduced
the following definitions

\bea
\beta^A&=& xm_t^2+(1-x)m_A^2 \nonumber \\
\eta^A&=&(x+xy+y^2)m_t^2+(1-x-y)m_A^2-xyq^2 \\
\gamma^A&=&\frac{xm_t^2+(1-x)m_A^2}{x^2m_t^2+(1-x)m_A^2}\,\,\,.
\nonumber 
\eea

\noindent The couplings $\xi^A_V$ and $\xi^A_A$ are now defined to be

\be
\xi^A_V=\frac{1}{4}\xi^U_{tt}\left(\xi^U_{ct}-\xi^{U*}_{tc}\right)
\,\,\,\,\,\mbox{and}\,\,\,\,\,\xi^A_A=\frac{1}{4}\xi^U_{tt}
\left(\xi^U_{ct}+\xi^{U*}_{tc}\right)
\ee

\noindent while $\alpha_V=\alpha_V^{\gamma,Z,g}$ and
$\alpha_A=\alpha_A^{\gamma,Z,g}$ as in Eq.~(\ref{axialvector_coupl}).
In this set of results we find many points of difference with respect
to Ref.~\cite{lukesavage}.

\subsection*{Contribution from diagrams with both $h^0$ and $A^0$}

\bea
A_M&=&\alpha_M\int_0^1\int_0^{1-x} 
dxdy\left\{\left[\log\frac{\eta^{M_1}}{\mu^2}-
m_t^2\frac{(1-y-xy-x^2)}{\eta^{M_1}}\right]\xi^{M_1}_V\right.\nonumber\\
&+&\left.\left[\log\frac{\eta^{M_2}}{\mu^2}+m_t^2\frac{(1-2x-y+xy+x^2)}
{\eta^{M_2}}\right]\xi^{M_2}_A\right\}\nonumber\\
B_M&=&\alpha_M\int_0^1\int_0^{1-x} 
dxdy\left\{\left[-\log\frac{\eta^{M_1}}{\mu^2}+
m_t^2\frac{(1-y-xy-x^2)}{\eta^{M_1}}\right]\xi^{M_1}_A\right.\nonumber\\
&+&\left.\left[-\log\frac{\eta^{M_2}}{\mu^2}-m_t^2\frac{(1-2x-y+xy+x^2)}
{\eta^{M_2}}\right]\xi^{M_2}_V\right\}\\
C_M&=&\alpha_Mm_t^2\int_0^1\int_0^{1-x}
dxdy\left\{-\frac{(1-y-xy-x^2)}{\eta^{M_1}}\xi^{M_1}_V-
\frac{(1-2x-y+xy+x^2)}{\eta^{M_2}}\xi^{M_2}_A\right\}\nonumber\\
D_M&=&\alpha_Mm_t^2\int_0^1\int_0^{1-x} 
dxdy\left\{\frac{(1-y-xy-x^2)}{\eta^{M_1}}\xi^{M_1}_A-
\frac{(1-2x-y+xy+x^2)}{\eta^{M_2}}\xi^{M_2}_V\right\}\nonumber
\eea

\noindent where we have used the following definitions

\bea
\eta^{M_1}&=& (1-2x-y+xy+x^2)m_t^2+ym_A^2+xm_h^2-xyq^2 \nonumber\\
\eta^{M_2}&=& (1-2x-y+xy+x^2)m_t^2+xm_A^2+ym_h^2-xyq^2 \,\,\,.
\eea

\noindent The couplings are now expressed in terms of
$\xi^{M_1}_{V,A}$ and $\xi^{M_2}_{V,A}$, defined as

\bea
\xi^{M_1}_V&=&\frac{1}{4}\xi^U_{tt}\left(\xi^U_{ct}-\xi_{tc}^{U*}\right)
\,\,\,\,\,,\,\,\,\,\,\xi^{M_1}_A=\frac{1}{4}\xi^U_{tt}\left(\xi^U_{ct}+
\xi_{tc}^{U*}\right)\nonumber\\
\xi^{M_2}_V&=&\frac{1}{4}\xi^U_{tt}\left(\xi^U_{ct}+\xi_{tc}^{U*}\right)
\,\,\,\,\,,\,\,\,\,\,\xi^{M_2}_A=\frac{1}{4}\xi^U_{tt}\left(\xi^U_{ct}-
\xi_{tc}^{U*}\right)
\eea

\noindent while the couplings of the photon and of the $Z$ boson to the
neutral scalar and pseudoscalar are given respectively by

\be
\alpha_M^\gamma=0\,\,\,\,\,\,,\,\,\,\,\,\,\alpha_M^Z=\frac{g_{\sss
W}}{2\cos\theta_{\sss W}}\,\,\,\,\,\mbox{and}\,\,\,\,\,\alpha_M^g=0
\,\,\,.
\ee

\noindent Also in this set of results we find many points of
difference with respect to Ref. \cite{lukesavage}.

\subsection*{Contribution from $H^+$}

\bea
A_C&=&\xi^C\left\{-4a_R\int_0^1 dx(1-x)\log\beta^C+
4b_L\int_0^1\int_0^{1-x}
dxdy\left(1+\log\frac{\eta^{C_1}}{\mu^2}-\frac{xyq^2}{\eta^{C^1}}
\right)\right.\nonumber\\
&+& \left. 2J_C\int_0^1\int_0^{1-x} dxdy\left[\log\frac{\eta^{C_2}}
{\mu^2}-m_t^2\frac{y(1-x-y)}{\eta^{C_2}}\right]\right\}\nonumber\\
B_C&=&A_C\\
C_C&=&-\xi^Cm_t^2\int_o^1\int_0^{1-x} dxdy\,y(1-x-y)
\left(\frac{4b_L}{\eta^{C_1}}-\frac{2J_C}{\eta^{C_2}}\right)\nonumber\\
D_C&=&-C_C\nonumber
\eea

\noindent where the following definitions have been used

\bea
\eta^{C_1}&=&(1-x-y)(m_C^2-xm_t^2)-xyq^2 \\
\eta^{C_2}&=&(x+y)m_C^2-y(1-x-y)m_t^2-xyq^2 \,\,\,.\nonumber
\eea

\noindent Using Eq.~(\ref{charged}), we should define $\xi^C$ to be a
linear combination of many terms. However, if we follow the Cheng and
Sher ansatz in Eq.~(\ref{coupl_sher}) with $\lambda_{ij}\!\sim\!O(1)$,
there is only one term among all the possible ones which gives the
leading contribution, and therefore we define

\be
\xi^C=\frac{1}{4}\xi^U_{tt}\xi^U_{ct}\,\,\,.
\ee

\noindent If different assumptions on the couplings of
Eq.~(\ref{coupl_sher}) are made, this statement will need to be
modified. However, in most cases, the main analytical results are
still valid.

Finally the remaining couplings are given for the different
vector bosons respectively by

\bea
a_R^\gamma&=&\frac{1}{3}e\,\,\,,\,\,\, b_L^\gamma=-\frac{1}{6}e\,\,\,,
\,\,\, J_C^\gamma=e\\
a_R^Z&=&-\frac{g_{\sss W}s^2_{\sss W}}{3c_{\sss W}}\,\,\,,\,\,\,
b_L^Z=-\frac{g_{\sss W}}{4c_{\sss W}}\left(1-\frac{2}{3}s^2_
{\sss W}\right)\,\,\,,\,\,\,J_C^Z=\frac{g_{\sss W}}{2c_{\sss
W}}(1-2s^2_{\sss W})\nonumber\\
a_R^g&=&\frac{1}{2}g_s\,\,\,,\,\,\, b_L^g=\frac{1}{2}g_s\,\,\,,
\,\,\, J_C^g=0\,\,\,.\nonumber
\eea
\bigskip

In our calculation we further assume that the $\xi^{U,D}_{ij}$
couplings are real, as we are not concerned with any phase dependent
effect, and symmetric for sake of simplicity. This amount to set in
the previous expressions $\xi^{U*}_{tc}\!=\!\xi^U_{ct}$. 


\begin{thebibliography}{99}

\bibitem{eetc} D. Atwood, L. Reina and A. Soni, \prd{\bf 53}, 
1199 (1996); 

\bibitem{mumutc} D. Atwood, L. Reina and A. Soni, Phys.\ Rev.\ Lett.\
{\bf 75}, 3800 (1995).

\bibitem{sher} T.P. Cheng and M. Sher, \prd{\bf35}, 3484 (1987); D{\bf44},
1461 (1991); See also Ref.~\cite{antaramian}.

\bibitem{antaramian} A. Antaramian, L.J. Hall, and A. Rasin, \prl\ {\bf69}, 
1871 (1992). 

\bibitem{hall} L.J. Hall and S. Weinberg, \prd{\bf48}, R979 (1993). 

\bibitem{savage} M.J. Savage, \pl\ B{\bf266}, 135 (1991). 

\bibitem{lukesavage} M. Luke and M.J. Savage, \pl\ B{\bf307}, 387 (1993).

\bibitem{gim} S.L. Glashow, J. Iliopoulos and L. Maiani, \prd{\bf 2},
1285 (1970).

\bibitem{glash} S. Glashow and S. Weinberg, \prd{\bf15}, 1958 (1977). 

\bibitem{hou} W.S. Hou, \pl\ B{\bf296}, 179 (1992); D. Chang, W.S. Hou
and W.Y. Keung; Phys.\ Rev.\ D{\bf48}, 217 (1993). 

\bibitem{hunter} For a review see J. Gunion, H. Haber, G. Kane,  and S.
Dawson, {\it The Higgs Hunter's Guide}, (Addison-Wesley, New York,
1990).  

\bibitem{knowles} C.D. Froggatt, R.G. Moorhouse and I.G. Knowles,
\np\ B{\bf 386}, 63 (1992). 

\bibitem{model3only} The only other contributions to the effective
$Ztc$ and $\gamma tc$ vertices are from the SM $W$-boson which we
are ignoring since we expect the contributions from model~III to be
much bigger due to the presence of both neutral and charged scalars
with large couplings to the top quark (see
Fig.~\ref{Zgtc_1loop}). Moreover, we expect model~III to give bigger
predictions, even with respect to model~I and model~II, due to the
presence of the new FC couplings proportional to $m_t$ and to the
particular structure of the charged scalar couplings (see
Eq.~(\ref{charged})). See Ref.~\cite{chang} for the explicit
calculation of $e^+e^-\rightarrow\bar t c +t\bar c$ in the SM and in
model I and model II 2HDM's. Our expectations are also confirmed by
the case of rare top decays, discussed in Sec.~\ref{tcgZ}, in which
the same form factors are used (see in particular Table~\ref{raretop}
for a quantitative comparison of the SM and of the different types
of 2HDM's).

\bibitem{chang} C.-H. Chang, X.-Q. Li, J.-X. Wang and M.-Z. Yang,
\pl\ B{\bf 313}, 389 (1993).

\bibitem{soni} G. Eilam, J.L. Hewett and A. Soni, \prd{\bf 44}, 1473
(1991). 

\bibitem{gunion} B. Grzadkowski, J.F. Gunion and P. Krawcyzk, \pl\
B{\bf 268}, 106 (1991).

\bibitem{16pi} We also observe that in Ref.~\cite{lukesavage} the form
factors used in the analytical expressions for the rates (see their
Eq.~(10)) should include the factor $1/(16\pi^2)$, which is factored
out in their Eq.~(9), where the form factors are defined.

\bibitem{csi} Note that in Fig.~2 of Ref.~\cite{lukesavage} the
authors factor out the FC coupling dependence: $|\xi_{ct}\xi_{tt}|^2$.
For $m_t\!\simeq\!180$ GeV and $m_c\!\simeq\!1.5$ GeV, taking as we
did $\lambda\!=\!1$, this amounts to multiplying their results by a
factor $\simeq 0.038$. Also note that, unlike them, we have not
multiplied the $Br(t\rightarrow Zc)$ by the
$Br(Z\rightarrow\mbox{leptons})\simeq 0.067$.

\bibitem{cline} D. Cline, Nucl.\ Instr.\ \& Meth.\ A{\bf350}, 24 (1994).

\bibitem{neuffer} D.V. Neuffer, {\it ibid}, p.~27.

\bibitem{barletta} W.A. Barletta and A.M. Sessler, {\it ibid}, p.~36;
A.G. Ruggiero, {\it ibid}, p.~45; S. Chattopadhyay {\it et al}., {\it
ibid}, p.~53.

\bibitem{palmer} R.B. Palmer, preprint, ``High Frequency $\mu^+ \mu^-$
Collider Design,'' SLAC-AAS-Note81, 1993.

\bibitem{barger} V. Barger, M. Berger, K. Fujii, J.F. Gunion, T. Han, C.
Heusch, W. Hong, S.K. Oh, Z. Parsa, S. Rajpoot, R. Thun and B. Willis,
Working group report, {\it Proc.\ of the First Workshop on the Physics
Potential and Development of $\mu^+\mu^-$ Colliders, Sausalito, CA},
preprint MAD-PH-95-873 (hep-ph/9503258).

\bibitem{bargertwo} V. Barger, M. Berger, J. Gunion, and T. Han, Phys.\ Rev.\
Lett.\ {\bf75}, 1462 (1995). 

\bibitem{mumu} D. Atwood and A. Soni, Phys.\ Rev.\ D{\bf52}, 6271 (1995).

\bibitem{inamilim} T. Inami and C.S. Lim, Prog.\ Th.\ Phys.\ {\bf 65},
297 (1981); Erratum, ibidem, 1772.

\bibitem{bpar} See e.g.\ A. Soni, Nucl.\ Phys.\ B(Proc.\ Suppl.) {\bf47},
43 (1996); G. Martinelli, ibid {\bf42}, 127 (1995).

\bibitem{shanker} B. Mc Williams and O. Shanker, \prd{\bf11}, 2853
(1980). See also, G. Beall, M. Bander and A. Soni, Phys.\ Rev.\ Lett.\
{\bf48}, 848 (1982).

\bibitem{ff} See Ref.~{\protect\cite{bpar}}.

\bibitem{buras_bb} A. J. Buras, M. Jamin and P.H. Weisz, 
\np\ B{\bf 347}, 491 (1990).

\bibitem{nierste} S. Herrlich and U. Nierste, \np\ B{\bf 419}, 292
(1994); \prd{\bf52}, 6505 (1995).

\bibitem{qcd} In fact, the SM predictions for the different
$F^0\!-\!\bar F^0$ mixings reported in Table~\ref{f0f0} include also
QCD corrections.

\bibitem{eps} Results presented at the {\it International Europhysics
Conference on High Energy Physics}, Brussels, 1995 and at the {\it
17th International Symposium on Lepton-Photon Interactions}, Beijing,
China, 1995. See also: LEP Electroweak Working Group Report 95-02. 

\bibitem{pdg} {\it Review of Particles Properties}, \prd{\bf50}
(1994). 

\bibitem{ohl} T. Ohl, G. Ricciardi and E.H. Simmons, \np\ B{\bf 403},
605 (1993).

\bibitem{grant} A.K. Grant, \prd{\bf 51}, 207 (1995).

\bibitem{alam} R. Ammar \etal CLEO Collaboration, \prl\ {\bf 71}, 674
(1993); M.S. Alam \etal CLEO Collaboration, \prl\ {\bf 74}, 2885
(1995).

\bibitem{moriond} Results presented at the {\it XXXIst Rencontres de
Moriond, ``Electoweak Interactions and Unified Theories''}, Les Arcs,
France, March 1996.

\bibitem{rb} The value of $R_b^{\rm expt}$ reported in
Eq.~(\ref{rb_expSM}) corresponds to the experimental measurement
obtained for $R_c=R_c^{\rm SM}=0.1724$. In fact, the experimental
measurement of $R_c$, $R_c^{\rm expt}=0.1598\pm 0.0070$ differs from
the SM prediction by about $1.8\sigma$ (see Ref.~\cite{moriond}).
However, due to the large error still present in this preliminary
measurement, we will not consider $R_c$ as a constraint for
model~III\null. In principle $R_c$ could play a very important role in
constraining the top quark FC couplings in model~III or in a more
general approach as well \cite{rbrc}. Therefore a better experimental
determination of $R_c$ is strongly advocated.

\bibitem{bamert} P. Bamert, C.P. Burgess, J.M. Cline, D. London and
E. Nardi, hep-ph/9602438, to appear in \prd.

\bibitem{rbrc} D. Atwood, L. Reina and A. Soni, hep-ph/9603210,
to appear in \prd.

\bibitem{isi} I. Dunietz, FERMILAB-PUB-96-104-T, hep-ph/9606247;
I. Dunietz, J. Incandela, F.D. Snider, K. Tesima and I. Watanabe,
FERMILAB-PUB-96-026-T, hep-ph/9606327 .

\bibitem{ciu_bsg1} M. Ciuchini, E. Franco, G. Martinelli, L. Reina and
L.  Silvestrini, \pl\ B{\bf 316}, 127 (1993); \np\ B{\bf 421}, 41
(1994).

\bibitem{buras_bsg} A.J. Buras, M. Misiak, M. M\"unz and S. Pokorski, 
\np\ B{\bf 424}, 137 (1994).  

\bibitem{ciu_bsg2} M. Ciuchini, E. Franco, G. Martinelli, L. Reina and L.
Silvestrini, \pl\ B{\bf 344}, 137 (1994).

\bibitem{bsg} According to the results of Sec.~\ref{mixing}, we neglect
the mixing with the first family. 

\bibitem{bsd} In passing, we should also mention that, we have also
examined the decay $b\rightarrow d\gamma$ in model~III. We found that
due to the different couplings involved, this decay is never
consistently enhanced with respect to the SM prediction.

\bibitem{massless} In our calculation all the quarks except the top
quark are taken to be massless, keeping $m_q\ne0$ only in the
$\xi^{U,D}_{ij}$ couplings as in Eq.~(\ref{coupl_sher}).

\bibitem{pich} J. Bernab\'eu, A. Pich and A. Santamaria, \np\ B{\bf 363},
326 (1991).

\bibitem{kuhn} A. Denner, R.J. Guth, W. Hollik and J.H. K\"uhn,
\zp\ C{\bf 51}, 695 (1991).

\bibitem{rurdrs} We assume that the changes for $R_u$, $R_d$ and $R_s$
in model~III with suppressed FC couplings for the first family,
i.e.\ in case~2 and case~3, are negligible.

\bibitem{lang} P. Langacker, hep-ph/9412361. To be published in ``{\it
Precision Tests of the Standard Electroweak Model}'', ed.\ by
P. Langacker, (World Scientific 1994).

\bibitem{cdf} F. Abe \etal [CDF], \prl\ {\bf74}, 2626 (1995);
S. Abachi \etal [$D\emptyset$], \prl\ {\bf74}, 2632 (1995).

\bibitem{bert} S. Bertolini, \np\ B{\bf 272}, 77 (1986).

\bibitem{2sigma} Due to the qualitative character of our analysis, at
this point it suffices to seek consistency with the experiment at the
$2\sigma$-level. Indeed, we took as reference the SM calculation
\cite{ciu_bsg2}, which is already affected by a large uncertainty, and
computed only the leading corrections due to the new scalar bosons of
model~III, i.e.\ without considering the complete leading order (LO)
effective hamiltonian analysis. From Fig.~\ref{bsg} we also note that,
for $M_c\ge 600$ GeV, model~III is difficult to distinguish from the
SM (again within $2\sigma$), unless the present SM calculation
($Br(B\rightarrow X_s\gamma)= (1.9\pm 0.6)\times 10^{-4}$
\cite{ciu_bsg2}) is improved. See also Ref.~\cite{bsg_th}.

\bibitem{bsg_th} The theoretical prediction of $Br(B\rightarrow
X_s\gamma)$ from Ref.~\cite{ciu_bsg2} includes some NLO QCD
corrections and the result could change in the future by a complete
NLO analysis. We could have used in our analysis the fully consistent
LO result for $Br(B\rightarrow X_s\gamma)$, which is a little higher (see
Ref.~\cite{buras_bsg,ciu_bsg2}), but this would not modify the
qualitative results that we are giving.

\bibitem{aleph} D. Buskulic \etal ALEPH Collaboration, CERN-PPE/96-30.

\bibitem{ua1_bs} UA1 Collaboration, C. Albajar \etal \pl\ B{\bf 262}, 163 
(1991).

\bibitem{cdf_bs} CDF Collaboration, F. Abe \etal
FERMILAB-PUB-96/040-E.

\bibitem{ua1} UA1 Collaboration, C. Albajar \etal \pl\ B{\bf
262}, 163 (1991).

\bibitem{grinstein} B. Grinstein, M.J. Savage and M.B. Wise, \np\ B{\bf
319}, 271 (1989).

\bibitem{buras_bsll} A.J. Buras and M. M\"unz, \prd{\bf 52},186 (1995).

\bibitem{poles} Even if the pole terms are not explicitly given in
Ref.~\cite{lukesavage}, they can be easily deduced from the
corresponding logarithmic term ($\log\mu^2$) in each diagram. 

\end{thebibliography}
\end{document}